\documentclass[12pt,preprint]{aastex}
\usepackage{amsmath}
\usepackage{amssymb}

\newcounter{qub}
\newcommand{\MC}{\multicolumn}

\shorttitle{{Spectrophotometry of Sextans A and B}}
\shortauthors{Kniazev et al.}

\begin{document}

\title{
Spectrophotometry of Sextans A and B: Chemical Abundances of H\,{\sc ii} Regions and
Planetary Nebulae\altaffilmark{0}
}

\author{Alexei Y.\ Kniazev\altaffilmark{1,2,4},
Eva K.\ Grebel\altaffilmark{3,1},
Simon A.\ Pustilnik\altaffilmark{4,5},
Alexander G.\ Pramskij\altaffilmark{4,5},
Daniel B.\ Zucker\altaffilmark{1}
}

\email{akniazev@eso.org, grebel@astro.unibas.ch, sap@sao.ru, pramsky@sao.ru,
       zucker@mpia.de}

\altaffiltext{0}{
Based on observations obtained at the European Southern Observatory,
La Silla, Chile (072.A-0087(B)).
}
\altaffiltext{1}{Max-Planck-Institut f\"{u}r Astronomie, K\"{o}nigstuhl 17,
      D-69117 Heidelberg, Germany}
\altaffiltext{2}{Present address: European Southern Observatory,
Karl-Schwarzschild-Strasse 2, 85748 Garching, Germany}
\altaffiltext{3}{Astronomical Institute of the University of Basel,
      Venusstrasse 7, CH-4102 Binningen, Switzerland}
\altaffiltext{4}{Special Astrophysical Observatory, Nizhnij Arkhyz,
      Karachai-Circassia, 369167, Russia}
\altaffiltext{5}{Isaac Newton Institute of Chile, SAO Branch}

\begin{abstract}
We present the results of high quality long-slit spectroscopy 
of planetary nebulae (PNe) and H\,{\sc ii} regions in the two dwarf 
irregular galaxies Sextans A and B, which belong to a small group of
galaxies just outside the Local Group.  The observations were obtained
with the ESO New Technology Telescope (NTT) multi-mode instrument (EMMI).
In Sextans A we obtained the element abundances
in its only known PN and in three H\,{\sc ii} regions with the classical
T$_{\rm e}$ method. The oxygen abundances in these three H\,{\sc ii} 
regions of Sextans A
are all consistent within the individual rms uncertainties, with the average
12+$\log$(O/H) = 7.54$\pm$0.06. The oxygen abundance of the PN in Sextans A
is, however, significantly higher: 12+$\log$(O/H) = 8.02$\pm$0.05. This PN is
even more enriched in nitrogen and helium, suggesting a classification
as a PN of Type~I.
The PN abundances of S and Ar, which are presumably unaffected by
nucleosynthesis in the progenitor star, are well below those
in the H\,{\sc ii} regions, indicating lower metallicity at the epoch of
the PN progenitor formation
($\sim$1.5 Gyr ago, according to our estimates based on the PN parameters).
In Sextans B we obtained spectra of one PN and six H\,{\sc ii}
regions. Element abundances with the T$_{\rm e}$ method could be derived
for the PN and three of the H\,{\sc ii} regions. 
For two of these H\,{\sc ii} regions, which have a separation of
only $\sim$70 pc in projection, the oxygen
abundances do not differ within the rms uncertainties, with a mean of
12+$\log$(O/H) = 7.53$\pm$0.05. The third H\,{\sc ii} region,
which is about 0.6 kpc northeast from the former two, is twice as metal-rich,
with 12+$\log$(O/H) = 7.84$\pm$0.05.
This suggests considerable inhomogeneity in the present-day metallicity
distribution in Sextans B.  Whether this implies a general chemical 
inhomogeneity among populations of comparable age in Sextans B, and thus
a metallicity spread at a given age, or whether we happen to see the 
short-lived effects of freshly ejected nucleosynthesis products prior
to their dispersal and mixing with the ambient interstellar medium
will require further study.
For the PN we measured an O/H ratio of 12+$\log$(O/H)=7.47$\pm$0.16,
consistent with that of the low-metallicity H\,{\sc ii} regions.
We discuss the new metallicity data for the H\,{\sc ii} regions and PNe 
in the context of the published star formation histories and published
abundances of the two dwarf irregular (dIrr) galaxies.  Both dIrrs
show generally similar star formation histories in the sense of
continuous star formation with amplitude variations, but differ in their
detailed enrichment time scales and star formation rates as a function
of time.  If we combine the photometrically derived estimates for the 
mean metallicity of the old red giant branch population in both dIrrs
with the present-day metallicity of the  H\,{\sc ii} regions, both
dIrrs have experienced chemical enrichment by at least 0.8 dex
(lower limit) throughout their history.
\end{abstract}


\section{Introduction}

The galaxies of the Local Group (LG) and its immediate surroundings are of
particular interest since their proximity permits us to study many aspects
of galaxy evolution in great detail.
The galaxy census of the LG and its surroundings is still incomplete.
A number of new nearby dwarf galaxy candidates were discovered in
recent years and have largely been confirmed
\citep[e.g.,][]{KarKara99,Armandroff99,GreGuh99,Whiting99,Ibata01,
Newberg02,Morr03,Yanny03,Zucker04a,Zucker04b}.
For recent discussions and reviews of the properties of galaxies in the 
Local Group and its surroundings, see \citet{Gre97,Gre99,Mateo98,GGH03}.
A recent review of the evolution and
chemistry of dwarf irregular galaxies is given in \citet{Gre04}.

As the closest {\em bona fide} dwarf group, the Antlia-Sextans group
\citep{Bergh99,Tul02} may hold
clues to the star formation histories and chemical evolution of galaxies
that presumably experienced little interaction and that evolved in relative
isolation.  In the present study, which is part of a larger effort to gain
a comprehensive picture of dwarf galaxy evolution in different environments 
\citep{Gresnap00}, we measure nebular abundances in two of
the dIrr galaxies in Antlia-Sextans, namely Sextans A and B
(see Table~\ref{tbl:General} for basic properties of both galaxies).
We determine nebular abundances both for H\,{\sc ii} regions and for
planetary nebulae (PNe) in these galaxies. The most massive, most luminous
galaxy in Antlia-Sextans, the dIrr NGC\,3109,
will not be discussed here; for its nebular abundances,
we refer to \citet{Lee03a} and \citet{KGPP04} and references therein.
With the telescopes presently available, spectroscopic 
abundance determinations for individual stars can only be carried out 
for luminous, massive young supergiants beyond the LG, 
so stellar abundance estimates
for older stellar populations have to rely on photometry (see Grebel,
Harbeck, \& Gallagher 2003 for results).  Emission-line
spectroscopy, on the other hand, can already be obtained with medium-sized
telescopes.  Emission-line spectra of H\,{\sc ii} regions provide a 
convenient means of obtaining present-day gas phase abundances in 
different areas of a galaxy.  Emission-line spectroscopy of planetary 
nebulae makes abundances of older populations (from a few 100 million
to many billion years) accessible. 

Sextans A  is at a distance of 1.32 Mpc \citep{Dolphin03a}.  Deep HST
imaging permitted photometric derivations of the recent and 
intermediate-age star formation and enrichment histories, but for ages of
several Gyr the uncertainties become large  \citep{Dolphin03b}.
The best fit is obtained for the assumption of a roughly constant metallicity
of $-$1.4 dex (with an rms scatter of 0.15 dex at all ages) throughout the
lifetime of Sextans A \citep{Dolphin03b}.  As is typical for dIrr galaxies
\citep{HunGal85,Tos91,Greggio93}, star formation
amplitude variations of factors of two to three occurred in the history
of Sextans A.  The star formation rate inferred for Sextans A
reveals recent increases starting a few Gyr ago \citep{Dolphin03b},
whereas the galaxy appears to have been more quiescent at earlier times.  
Spectroscopic abundance determinations for three A supergiants yield
$\langle$[Fe/H]$\rangle = -1.03 \pm 0.1$ \citep{Kaufer04}.
The heavy element gas abundance for Sextans A
was first determined by \citet[][SKH89 hereafter]{SKH89}.
They presented only an empirical
estimate of O/H ($12+$ log(O/H)$ = 7.49$), which has a characteristic rms 
uncertainty of $\sim$50\%.
One of the goals of our current study is to re-determine nebular abundances with
higher accuracy.

Sextans B is a dIrr at a distance of 1.36 Mpc \citep{Kar02}.  Like in Sextans
A, the star formation history has likely been complex, with the long periods
of relatively low star formation rates, punctuated by periods of short
star-forming bursts \citep{Tos91,Sak97}.
As far as its global properties are concerned, Sextans B
is considered to be a ``twin'' of Sextans A, but Sextans B is of particular
interest because the previous measurements of the nebular oxygen abundance 
for this galaxy by \citet{SCV96}, SKH89 and \citet[][MAM90 hereafter]{MAM90}
differ considerably.
Here we re-measure the H\,{\sc ii} region oxygen
abundances with additional high quality spectroscopy and check whether
the apparent discrepancy remains, and compare these abundances with
the abundance of the candidate planetary nebulae (PNe) found by \citet{Magr02}.

The contents of this paper are organized as follows.
\S~\ref{txt:Obs_and_Red} gives the description of all observations and
data reduction.
In \S~\ref{txt:results} we present our results, 
and discuss them in \S~\ref{txt:disc}.
The conclusions drawn from this study are summarized in
\S~\ref{txt:summ}.
For the remainder of this paper, we adopt the revised solar value
of the oxygen abundance, 12+$\log$(O/H) = 8.66 \citep{Asp04}.

\section{Observations and reduction}
\label{txt:Obs_and_Red}

\subsection{NTT Observations}

Spectrophotometric observations of H\,{\sc ii} regions and PNe in the
dwarf irregular galaxies Sextans A and Sextans B
were conducted with the New Technology Telescope (NTT), 
a 3.5-meter telescope at the European Southern Observatory (ESO), La Silla,
on the nights of 2004 February 12 -- 14.
All spectral observations were made with the
low-resolution long-slit spectroscopy (RILD) mode of the EMMI
(ESO Multi-Mode Instrument)
used in conjunction with two mosaic 2048$\times$4096 CCD detectors.
With a pixel scale of 0\farcs166 the effective field of view was
9\farcm9$\times$9\farcm1.
Using a binning factor of 2, our final spatial sampling was 
0\farcs33 pixel$^{-1}$.
We used a $8\arcmin \times 2\arcsec$ slit for all spectral observations.
Grisms \#5 (spectral range 3800 -- 7000 \AA,
a final scale perpendicular to the slit of 1.63 \AA\ pixel$^{-1}$
and a spectral resolution (FWHM) of $\sim$7 \AA)
and \#6 (spectral range 5750 -- 8670 \AA,
a final scale perpendicular to the slit of 1.44 \AA\ pixel$^{-1}$
and a FWHM of $\sim$6.5 \AA)
were used to cover the spectral range with all the lines of interest.
The seeing during the observations was very stable and varied from night
to night in the range of 0\farcs4 to 0\farcs6.
Total exposure times were 60 or 90 minutes for each grism.
Each exposure was broken up into 2--3
subexposures, 30 minutes each, to allow for removal of cosmic rays.
Spectra of He--Ar comparison arcs were
obtained to calibrate the wavelength scale. The
spectrophotometric standard stars Feige~56, GD~108, GD~50 and G~60-54
\citep{Oke90,Bohlin96} were observed with a 10\arcsec\ slit width at the beginning, middle
and end of each night for flux calibration.

\subsection{The NTT pointing strategy and target nomenclature}

One or two-minute H$\alpha$ acquisition images were obtained before the spectroscopic
observations in order to select an optimal position of the slit. We tried to place
the slit over the knots with maximal H$\alpha$ flux to increase the probability of
detecting the [O\,{\sc iii}] $\lambda$4363 emission line, and thus being able to
calculate the electron temperature directly.

\begin{figure}
    \begin{center}
    \epsscale{1.0}
    \includegraphics[clip=,angle=0,width=14.0cm]{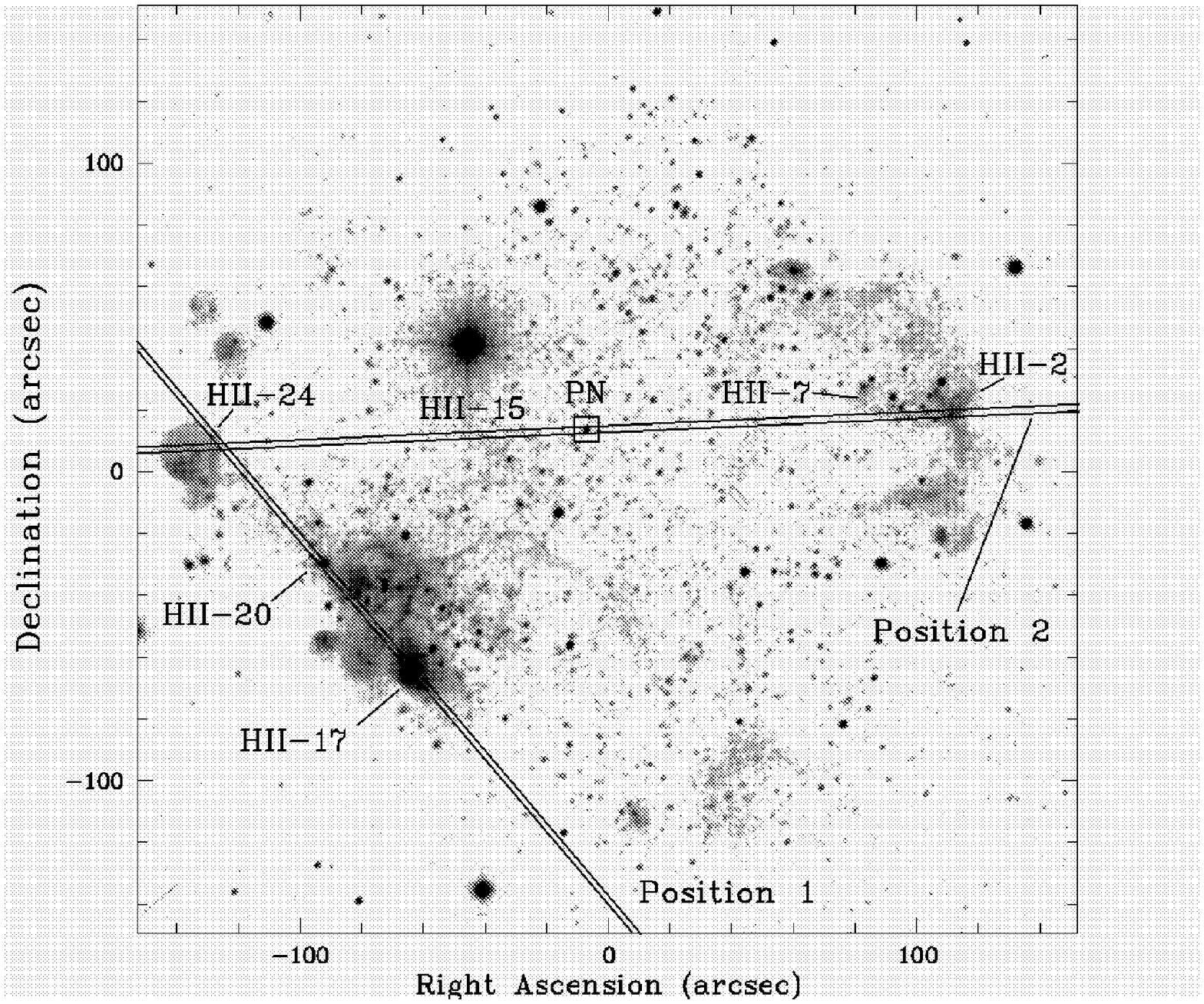}
    \caption{
     The combined H$\alpha$ image of Sextans A, obtained with the ESO NTT.
     North is up and East is to the left.
     The contrast is adjusted to highlight both bright and faint H$\alpha$
     emission.
     Black objects on the image indicate bright sources.
     Long-slit orientations are shown; the slit width is 2\arcsec.
     The position of the PN is marked by a square.
     Only those H\,{\sc ii} regions from \citet{HKS94} through which our slit positions passed are labeled.
     At the adopted distance of 1.32 Mpc, 1\arcsec\ = 6.4 pc.
    \label{fig:SexA_Ha}}
    \end{center}
\end{figure}

\begin{figure}
    \begin{center}
    \epsscale{1.0}
    \includegraphics[clip=,angle=0,width=15.0cm]{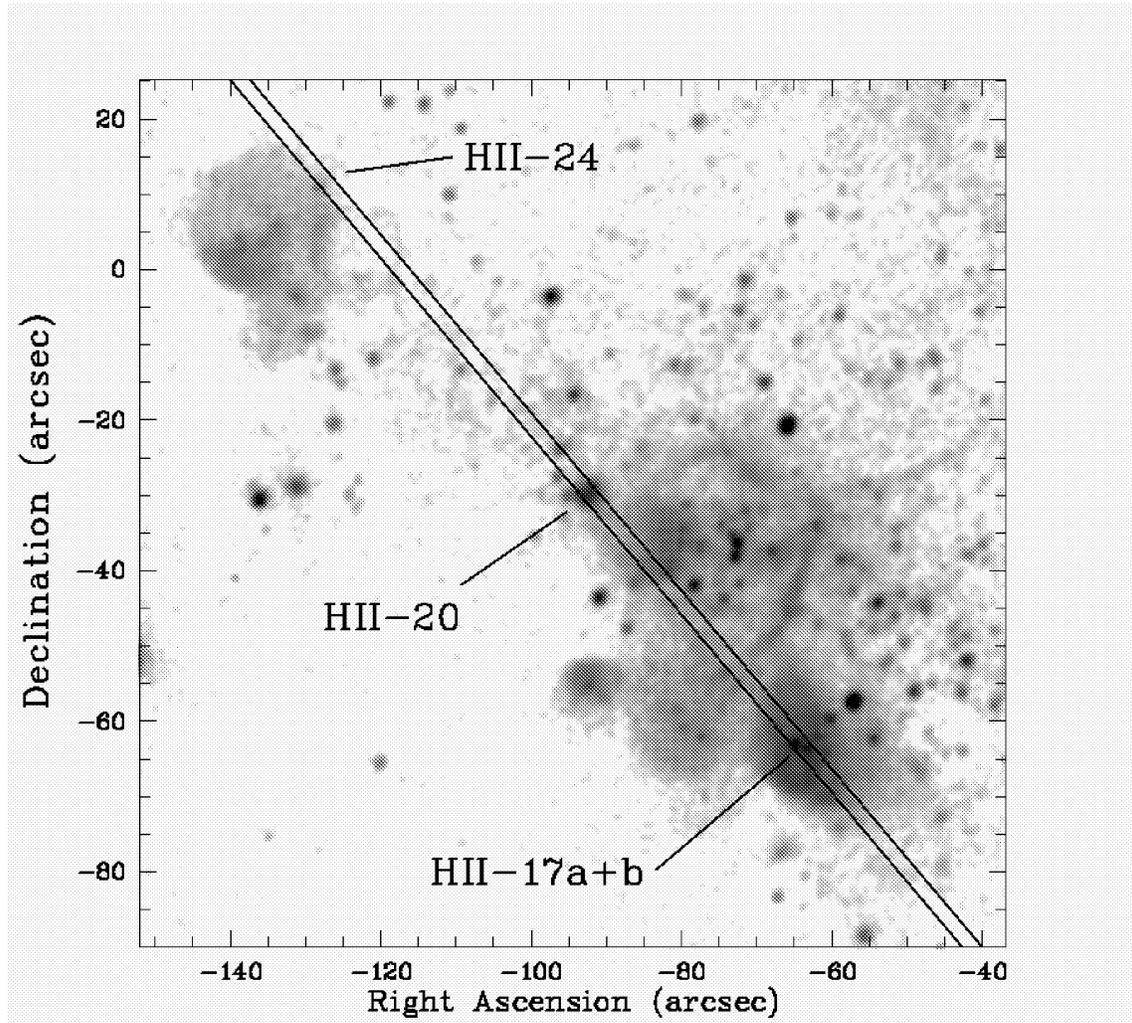}
    \caption{
     Enlargement of a portion of the combined H$\alpha$ image of Sextans A
     showing the H\,{\sc ii} regions 24, 20, and 17 in more detail.
     The position of the 2\arcsec\ width slit is marked. The H$\alpha$ image
     is rescaled to show the internal structure of H\,{\sc ii}$-$17.
     Two compact bright knots, located on the rim of a circular shell,
     fall well inside the slit. The other shells are well resolved in
     the areas between H\,{\sc ii}$-$20 and
     H\,{\sc ii}$-$17, where H\,{\sc ii}$-$16 18 and 19 \citep{HKS94} are
     located,
     and at the edge of H\,{\sc ii}$-$24, where the slit is positioned.
    \label{fig:SexA_Ha_b}}
    \end{center}
\end{figure}

For Sextans~A we use the H\,{\sc ii} region nomenclature
published by \citet{HKS94}.
One of our slits crossed the extended regions H\,{\sc ii}$-$17 and 19 and
the compact region H\,{\sc ii}$-$20 (see Figure~\ref{fig:SexA_Ha} for 
more details).  These H\,{\sc ii} regions correspond to the regions 1, 3, 
and 2 of SKH89.  Additionally, this slit position crossed the extended
regions H\,{\sc ii}$-$18, 22, and 24 \citep{HKS94}.  Thanks to the good
seeing and resulting high angular resolution, $\sim$0\farcs5, H\,{\sc ii}$-$17 was 
well resolved into two compact bright knots with an angular separation of
$\sim$1\farcs5 (called hereafter H\,{\sc ii}$-$17a and 17b), which seem 
to be located at the intersection of two shell-like structures 
(Figure~\ref{fig:SexA_Ha_b}).  The second slit position went through the 
PN.  The coordinates of the PN are given in \citet{Magr03}, and a finding
chart was published in \citet{JL81}.  This PN is called H\,{\sc ii}$-$13 
in \citet{HKS94}.  The second slit position also crosses H\,{\sc ii}$-$2 
and 7 in the West, H\,{\sc ii}$-$15 in the central part of the galaxy, and
the northern edge of H\,{\sc ii}$-$24 in the East (see also 
Figure~\ref{fig:SexA_Ha}).

In Sextans~B we adopt the H\,{\sc ii} region nomenclature suggested by 
\citet{SHK91}. For the observed PN candidate in this galaxy we use the
nomenclature of \citet{Magr02}, who named it PN3.
One slit position went through
H\,{\sc ii}$-$1, 4, 5, and the edge of 2, and the other one crossed 
H\,{\sc ii}$-$2, 6, and 11 (see Figure~\ref{fig:SexB_Ha} for more details).

\subsection{Data reduction}

The two-dimensional spectra were bias subtracted and flat-field corrected
using IRAF\footnote{IRAF: the Image Reduction and Analysis Facility is
distributed by the National Optical Astronomy Observatory, which is
operated by the Association of Universities for Research in Astronomy,
In. (AURA) under cooperative agreement with the National Science
Foundation (NSF).}.
Cosmic ray removal was done with FILTER/COSMIC task in
MIDAS.\footnote{MIDAS is an acronym for the European Southern
Observatory package -- Munich Image Data Analysis System.}
We use the IRAF software routines IDENTIFY, REIDENTIFY, FITCOORD, TRANSFORM to
perform the wavelength calibration and to correct each frame for distortion 
and tilt. To derive the sensitivity curve, we
fitted the observed spectral energy distribution of the standard stars
with a high-order polynomial.
All sensitivity curves observed during each night were compared and we
found the final curves to have a precision better than 2\% over the
whole optical range, except for the region blueward of
$\lambda$4000 where the sensitivity drops off rapidly.
All two-dimensional spectra obtained with the same grism for the same object
were averaged.
One-dimensional (1D) spectra were then extracted from averaged frames
using the IRAF APALL routine.
For the H\,{\sc ii}-regions 1D spectra were extracted from areas along the slit, where
\mbox{$I$($\lambda$4363~\AA)} $>$ 1.5$\sigma$ ($\sigma$ is the dispersion of
the noise statistics around this line).
For the PNe, 1D spectra were extracted to get the total light.

The resulting reduced and extracted spectra of the PNe in
Sextans~A and Sextans~B obtained with grism \#5
are shown in Fig.~\ref{fig:SexA_PN_spec}, and 
part of the reduced and extracted spectrum of the PN in Sextans~A
obtained with grism \#6 is shown in Fig.~\ref{fig:SexA_PN_spec_red}.
The reduced and extracted 1D spectra of H\,{\sc ii} regions in Sextans A,
obtained with grism \#5 and in which the [O~{\sc iii}] 
$\lambda$4363 line was detected, are shown in Fig.~\ref{fig:SexA_HII_spec}.
Fig.~\ref{fig:SexB_HII_spec} shows the reduced and extracted 1D spectra obtained with grism \#5 of H\,{\sc ii} 
regions in Sextans~B in which the [O~{\sc iii}] $\lambda$4363 line was
detected.

After 1D-spectra were extracted, we used a method for measuring 
emission-line intensities described in detail in 
\citet{Kniazev00a,Pust03,Kniazev04}:
(1) the software is based on the MIDAS Command Language;
(2) the continuum noise estimation was done using
the absolute median deviation (AMD) estimator;
(3) the continuum was determined with the algorithm from \citet{Sh_Kn_Li_96};
(4) the programs dealing with the fitting
of emission/absorption line parameters are based on the MIDAS
{\tt FIT} package;
(5) every line was fitted with the Corrected Gauss-Newton method
as a single Gaussian superimposed on the continuum-subtracted spectrum.
Some close lines
were fitted simultaneously as a blend of two or more Gaussian features:
the H$\alpha$  $\lambda$6563 and [N~{\sc ii}] $\lambda\lambda$6548,6584 lines,
[S~{\sc ii}] $\lambda\lambda$6716,6731 lines, and
[O~{\sc ii}] $\lambda\lambda$7320,7330 lines;
(6) the final errors in the line intensities, $\sigma_{\rm tot}$, include
two components: $\sigma_{\rm p}$, due to the Poisson statistics of line photon flux,
and $\sigma_{\rm c}$, the error resulting from the creation of
the underlying continuum and calculated using the AMD estimator;
(7) all uncertainties were propagated in the
calculation of abundances and are accounted for in the
accuracies of the presented element abundances.

The emission lines He~{\sc i} $\lambda$5876, [O~{\sc i}] $\lambda$6300,
[S~{\sc iii}] $\lambda$6312, [O~{\sc i}] $\lambda$6364,
H$\alpha$  $\lambda$6563, [N~{\sc ii}] $\lambda\lambda$6548,6584,
He~{\sc i} $\lambda$6678, and [S~{\sc ii}] $\lambda\lambda$6716,6731
were usually detected independently in both grism spectra. Their
intensities and equivalent widths were averaged after applying weights
inversely proportional to the errors of the measurements.

\subsection{Physical conditions and determination of heavy element abundances}

The electron temperature, number densities, ionic and total element
abundances for oxygen were calculated in the same manner as
in \citet{Kniazev04}, where the
[O\,{\sc ii}] $\lambda$3727,3729 doublet was also not observed,
and O$^+$/H$^+$ was calculated using intensities of the
[O\,{\sc ii}] $\lambda$7320,7330 lines.
The contribution to the intensities of the [O\,{\sc ii}]
$\lambda$7320,7330 lines due to recombination was not taken into account
since this is negligible compare to other uncertainties (namely, $<1\%$ as
follows from \citet{Liu00}).
Total element abundances for Ne, N, S and Ar were derived after
correction for unseen stages of ionization
as described in \citet{ITL94} and \citet{TIL95}:
(1) the electron temperatures for different ions
were either calculated from the relations in \citet{Gar92}
or using the equations following
from the H\,{\sc ii} photoionization models of \citet{Stas90}:
%
%
\begin{equation}
t_e({\rm O\;II})=0.243+
t_e({\rm O\;III}) \left[ 1.031 - 0.184\,t_e({\rm O\;III}) \right],
						     \label{eq:tOII}
\end{equation}
\begin{equation}
t_e(NeIII)=t_e(OIII),                                \label{eq:tNeIII}
\end{equation}
\begin{equation}
t_e(NII)=t_e(OII),                                   \label{eq:tNII}
\end{equation}
\begin{equation}
t_e(SII)=t_e(OII),                                   \label{eq:tSII}
\end{equation}
\begin{equation}
t_e(ArIII)=0.83t_e(OIII)+0.17,                       \label{eq:tArIII}
\end{equation}
\begin{equation}
t_e(SIII)=t_e(ArIII),                                \label{eq:tSIII}
\end{equation}
\begin{equation}
t_e(ArIV)=t_e(OIII),                                \label{eq:tArIV}
\end{equation}
where t$_e$=T$_e$/10$^4$K.
(2) The ionization correction factors ICF(A) for different elements
were calculated using the equations from \citet{TPP77,Gar90,ITL94,TIL95}:
\begin{equation}
ICF(N) = \frac{O}{O^+},                             \label{eq:ICFN}
\end{equation}
\begin{equation}
ICF(Ne) = \frac{O}{O^{++}},                          \label{eq:ICFNe}
\end{equation}
\begin{equation}
ICF(S) = \frac{S}{S^++S^{++}} = \{0.013+x[5.10+x(-12.78 \nonumber
\end{equation}
\begin{equation}
 +x(14.77-6.11x))]\}^{-1},
\label{eq:ICFS}
\end{equation}
\begin{equation}
ICF(Ar)  =  \frac{Ar}{Ar^{++}+Ar^{+++}} \nonumber
\end{equation}
\begin{equation}
	 = \{0.99+x[0.091+x(-1.14+0.077x)]\}^{-1},
\label{eq:ICFAr1}
\end{equation}
\begin{equation}
ICF(Ar) = \frac{Ar}{Ar^{++}} = [0.15+x(2.39-2.64x)]^{-1},     \label{eq:ICFAr2}
\end{equation}
where $x$ = O$^+$/O. In those cases where the [Ar {\sc iv}] $\lambda$4740
line was not detected, equation~\ref{eq:ICFAr2} was used instead of
equation~\ref{eq:ICFAr1}.
The He abundance was calculated following the method
described
in detail by \citet{ITL97} and \citet{IT98b}.

The line [N\,{\sc ii}] $\lambda$5755 was detected in the spectrum of the
Sextans A PN, which allowed us to determine $t_e(NII)$ directly.
This was done in the same way as for $t_e(OIII)$ using the \citet{Aller84}
equation:
\begin{equation}
\frac{I(6584 + 6548)}{I(5755)} = C_T \Big[\frac{1+a_1 x}{1+a_2 x}\Big] 10^{1.086/t_{\rm e}}
\end{equation}
Here the constants $C_T$, $a_1$, and $a_2$ for different temperatures
were taken from \citet{Aller84} and interpolated using an 
iterative procedure. $x = 10^{-2} N_e T_e^{-1/2}$
is a term describing the dependence on the electron density $N_e$ and
$T_e$.

Additionally, a strong nebular He II $\lambda$4686 emission
line is detected in the spectrum of the Sextans~A PN,
implying the presence of a non-negligible amount of O$^{3+}$.
In this case its abundance was derived from the relation
taken from \citet{IT99}:
\begin{equation}
{\rm O}^{3+}=\frac{{\rm He}^{2+}}{{\rm He}^+}({\rm O}^++{\rm O}^{2+})
\label{eq:O3+}
\end{equation}
After that, the total oxygen abundance is equal to
\begin{equation}
{\rm O}={\rm O}^++{\rm O}^{2+}+{\rm O}^{3+}       \label{eq:O}
\end{equation}

\section{Results}
\label{txt:results}

\subsection{Physical conditions and chemical abundances in H\,{\sc ii} regions}
\label{txt:Chem_HII}

The observed emission line intensities $F(\lambda)$, and those corrected
for interstellar extinction and underlying stellar absorption $I(\lambda)$
are presented in Table~\ref{tbl:SextA_HII_int} for the three H\,{\sc ii} 
regions in Sextans A,
in which the [O\,{\sc iii}] $\lambda$4363 line was detected:
H\,{\sc ii}$-$17a, H\,{\sc ii}$-$17b and H\,{\sc ii}$-$20.
The H$\beta$ equivalent width $EW$(H$\beta$),
the absorption equivalent widths $EW$(abs) of the Balmer lines,
the H$\beta$ flux, and the extinction coefficient $C$(H$\beta$)
(this is a sum of the internal extinction in Sextans~A and foreground extinction
in the Milky Way) are also listed there.
In~Table~\ref{tbl:SextB_HII_int} we present similar data for the three
H\,{\sc ii} regions H\,{\sc ii}$-$1, 2 and 5 in Sextans~B.
The range of $C$(H$\beta$) measured in the H\,{\sc ii} regions of Sextans~A
corresponds
to values of B-band extinction A$_{\rm B}$ ranging from 0\fm20 to 0\fm54.
Accounting for the Milky Way foreground extinction of 0\fm19 in the direction 
to this galaxy \citep{Schlegel98} the internal extinction values are in 
the range of 0\fm01 to 0\fm35,
implying a low dust content in these H\,{\sc ii} regions.
For Sextans~B, the values of A$_{\rm B}$ calculated from the spectra 
for H\,{\sc ii}$-$5 and 1 are consistent to within the uncertainty
in the foreground extinction in the direction of this galaxy, 0\fm14.
The A$_{\rm B}$ for H\,{\sc ii}$-$2 corresponds to an additional
internal extinction of 0\fm35.

In Table~\ref{tbl:SextA_HII_ab} we give the derived physical
conditions and ion and element abundances for the above mentioned
three H\,{\sc ii} regions in Sextans~A. The corresponding data for three 
H\,{\sc ii} regions in Sextans~B are
presented in Table~\ref{tbl:SextB_HII_ab}. As evident from these Tables,
we detected the faint [O\,{\sc iii}] $\lambda$4363 line at the level of 4
to 8 $\sigma_{\rm noise}$ and the [O\,{\sc ii}] $\lambda$7325 line at the level
of 4 to 10 $\sigma_{\rm noise}$. This results in O/H uncertainties in
the individual H\,{\sc ii} regions on the order of 0.05 to 0.11 dex.
Since the flux uncertainties of faint lines of other ions are in general
higher, the resulting errors of N/H, S/H and Ar/H
in the individual H\,{\sc ii} regions lie in the range of 0.1 to 0.2 dex.
The line [S\,{\sc iii}] $\lambda$6312 is not detected in H\,{\sc ii}$-$17b
of Sextans~A, and the derived value of S/H should therefore be treated as a lower limit. Hence we do not list S/H for this H\,{\sc ii} region.
The line [Ne\,{\sc iii}] $\lambda$3868 is detected in only one H\,{\sc ii}
region, due to the significant drop in
sensitivity below $\sim$$\lambda$4000~\AA.

All three H\,{\sc ii} regions in Sextans~A from Table~\ref{tbl:SextA_HII_ab}
show oxygen abundances consistent with each other to within the uncertainties, and hence a weighted average value of 12+log(O/H)=7.54$\pm$0.06
can be derived for this galaxy based on our data.
For the two H\,{\sc ii} regions in Sextans~B
(H\,{\sc ii}$-$1 and H\,{\sc ii}$-$2 from Table~\ref{tbl:SextB_HII_ab}),
the oxygen abundances also are very similar and within their
rather small rms uncertainties can be treated as equal.
Their weighted mean O/H value corresponds to
12+$\log$(O/H)=7.53$\pm$0.05.

The average relative abundances N/O, S/O and Ar/O
for all three H\,{\sc ii} regions in Sextans~A
($-$1.58$\pm$0.07, $-$1.45$\pm$0.18, and $-$2.06$\pm$0.09 respectively)
and for two low-metallicity H\,{\sc ii} regions in Sextans~B
($-$1.53$\pm$0.07, $-$1.59$\pm$0.09, and $-$2.04$\pm$0.06)
are quite consistent with the average values for the most metal-poor
H\,{\sc ii} galaxies
obtained by \citet{IT99} and confirmed in many subsequent observations
\citep[e.g.,][]{Kniazev00a,Kniazev00b,Gus01,Gus03a,Gus03b,Gus03c,Pust03,Pust04},
implying production in massive stars.
The relative abundances of 
N/O, Ne/O, S/O and Ar/O for the third H\,{\sc ii} region
in Sextans~B analyzed here
are also consistent with the mean values determined by
\citet{IT99} for their high-metallicity subsample of H\,{\sc ii} galaxies.

\subsection{Physical conditions and elemental abundances in PNe}
\label{txt:Chem_PN}

We present the observed and Balmer line extinction corrected 
emission line intensities for the two observed PNe
in Table~\ref{tbl:SextAB_PN_int}.
Our  fluxes for the emission lines [O~{\sc iii}] $\lambda$5007
and H$\alpha$
in both PNe are consistent to within the observational uncertainties with
the fluxes measured by \citet{Magr03} and \citet{Magr02}.
The measured $C$(H$\beta$) in the Sextans~A and Sextans~B PNe corresponds
to A$_{\rm B}$ of 0\fm66 and 0\fm71, respectively
(A$_{\rm V}$ of 0\fm50 and 0\fm54). After
accounting for the Galactic contribution, this suggests 
an intrinsic extinction
A$_B$ of 0\fm47 and 0\fm57, which is higher than the values
for H\,{\sc ii} regions in these galaxies.
This result is consistent with the fact that the progenitors of PNe produce
additional dust in the late stages of their evolution.
In Table \ref{tbl:SextAB_PN_ab} we give the derived physical
conditions, ion and element abundances for both PNe.

For the PN in Sextans~A the quality of the spectral data is comparable
to the quality of the H\,{\sc ii} region spectra. The resulting
uncertainties in
He/H, O/H and N/H are at the level of 0.03--0.05 dex.
The values of O/H and N/H for the PN appear to be significantly
higher than observed in the H\,{\sc ii} regions of this galaxy
(by a factor of $\sim$3 and $\sim$250, respectively).
With $\log$(N/O) = 0.39 and He/H = 0.107$\pm$0.008,
according to the definition of \citet[][$\log$(N/O)$>-$0.1]{KB94}
or to the definition in \citet[][$\log$(N/O)$>-$0.6 and He/H $>$ 0.10]{LD96},
the PN in Sextans~A should be classified as a Type~I.

Since the line fluxes of PN3 in Sextans~B are about a factor of 2--3 
(apart [N\,{\sc ii}]) lower than for the Sextans~A PN,
the resulting signal-to-noise for the spectrum of PN3 in Sextans~B
is significantly less. Therefore, from this spectrum it was only possible
to derive values of O/H and N/H
with final uncertainties of 0.16 and 0.18 dex, respectively.
The observed properties of PN3 allow us to exclude a classification 
as a PN of Type~I.
However, the quality of the PN spectrum is not sufficient to perform
the detailed comparison of its abundances with those of the
H\,{\sc ii} regions in Sextans~B.

\subsection{Physical Parameters of observed PNe}
\label{txt:Par_PNe}

The self-consistent age-dating of PNe progenitors
requires the determination of the bolometric
luminosity and the effective temperature of the PN central star in
combination
with photoionization modelling and PN evolutionary tracks \citep{Dopita97}.
This is well beyond the scope of our paper.
However, based on our data, it is possible to make some rough
estimates, which we summarize in Table~\ref{tbl:PN_par}.
The effective temperatures (T$_{eff}$) of the PN progenitors can be
estimated by the method developed by \citet{Zan27}.
Since the method is applicable for the case of optically thick nebulae,
we have to first check whether our PNe are optically thick.
The criteria of this were suggested by \citet{Kaler83} on the results
of analysis of the dereddened I(3727)/I(H$\beta$), I(4686)/I(H$\beta$)
and I(6584)/I(H$\alpha$) line ratios
for the large sample of PNe covered a wide range of optical thickness.
According to these criteria,
the dereddened line ratios I(4686)/I(H$\beta$) and I(6584)/I(H$\alpha$)
that reflect the Zanstra temperature ratio TR=T$_z$(He\,{\sc ii})/T$_z$(H)
\citep{Kaler83}, show that our PNe are certainly optically thick, since:
(a) PN in Sextans~A has TR$\le$1.2 and is the Case~1 following
\citet{Kaler83} classification \citep[class a i from][]{Seaton66};
(b) PN3 in Sextans~B has TR$>$1.2 and I(4686)$<$0.9 and is the Case~2
following \citet{Kaler83} classification \citep[class a ii from][]{Seaton66}.
From the spectrum of PN3 in Sextans~B we estimated a 2$\sigma$ upper
limit for the flux of He\,{\sc ii} $\lambda$4686 line as 8\%
of I(H$\beta$).
After that, for the calculation of the Zanstra temperature we used the
equation suggested by \citet{KJ89}:
\begin{equation}
\log{T_{eff}} = 4.905 + 1.11162 \times 10^{-2} I_c (\lambda4686) \nonumber
\end{equation}
\begin{equation}
- 1.10692 \times 10^{-4} I_c^2 (\lambda4686)  \nonumber
\end{equation}
\begin{equation}
+ 6.20572 \times 10^{-7} I_c^3 (\lambda4686),
\end{equation}
where $I_c (\lambda4686)$ is the
flux of He\,{\sc ii} $\lambda$4686 in the units of I(H$\beta$)
with I(H$\beta$) = 100.

The total luminosities of the PN central stars were derived
from the relation given in \citet{GP89,ZP89},
using the H$\beta$ absolute fluxes, extinction $C$(H$\beta$) and
distances:
\begin{equation}
L = 150 \times L(H\beta)
\end{equation}
Masses were derived from
the theoretical evolutionary tracks
of \citet{VW94} for Z=0.001 (1/20 of Z$_{\odot}$). The ages were derived
using the evolutionary lifetimes of the various phases of the progenitor
stars from \citet{VW93}, also for Z=0.001.

In addition, since the PNe 1D spectra were extracted taking the total light
of a nebula, we synthesized the $VRI$ colors of the PNe
using zeropoints from the spectrum of Vega \citep{Cas_Kur94}.
These colors were transformed into the standard Johnson-Cousins $UBVRI$
system
as specified by \citet{Bessel90} and \citet{Bessel98}.
The $B$-band was not included in Table~\ref{tbl:PN_par} since,
due to the drop in sensitivity,
the spectra become very noisy below $\sim$$\lambda$4000~\AA.

\section{Discussion}
\label{txt:disc}

\subsection{PN abundances vs. H\,{\sc ii} region abundances}

H\,{\sc ii} region abundances mainly provide information about
$\alpha$-process elements, produced predominantly in short-lived massive stars.
In contrast to H\,{\sc ii} regions, some elemental abundances
in PNe are affected by nucleosynthesis in the PN progenitor stars.
It is well known that newly synthesized material can be dredged up
by convection in the envelope, significantly altering the abundances of He, C,
and N in the surface layers during the evolution of 1--8 M$_\odot$ stars on
the giant branch and asymptotic giant branch \citep[AGB;][]{LD96,BS99,HKB00}.
In addition, if during the thermally pulsing phase of AGB evolution
convection ``overshoots'' into the core, significant amounts of $^{16}$O
can be mixed into
the inter-shell region and may be convected to the surface \citep{Her00,Bl01}.
In combination, all of the above factors mean that only the Ne, S, and Ar 
abundances,
observed in both H\,{\sc ii} regions and PNe, can be considered as reliable
probes of the enrichment history of galaxies, unaffected by the immediately
preceding nucleosynthesis in the progenitor stars.
Therefore, in further discussions of PN progenitor metallicity we do not
use O/H, He/H, and N/H as reliable metallicity indicators
and base our conclusions
on chemical evolution only on PNe abundances of Ne, S, and Ar.

\subsection{Observed heavy element abundances}

\subsubsection{Sextans A abundances}
\label{txt:disc_abun_SextA}

The value of 12+log(O/H)=7.54$\pm$0.06, derived here as the average of
three H\,{\sc ii} regions in Sextans A, is in good agreement with the 
earlier derived value 12+log(O/H)=7.49$\pm$0.3 from SKH89.
However, the previous value was based
on spectra of four separate H\,{\sc ii} regions with only a marginal 
detection of [O\,{\sc iii}] $\lambda$4363 in one of them, such that SKH89
used a combined estimate for all four H\,{\sc ii} regions based on the 
strong-line empirical relation
of \citet{Pagel89}.  It is worth noting that the use of
modern  empirical calibrations of the strong-line relation
\citep{McGaugh91,Pilyugin01} for the published data from SKH89
gives 12+$\log$(O/H) in the range of 7.7 to 8.0.

Recently \citet{Kaufer04} measured the present-day stellar
metallicity of Sextans~A using
three A-type supergiants with ages of $\sim$10 Myr.
Their average $\alpha$-element abundance, derived from Mg~I lines,
relative to the solar value is
$<$[$\alpha$(Mg I)/H]$>$ = --1.09 $\pm$ 0.02 $\pm$0.19,
which corresponds to an oxygen abundance of 12+$\log$(O/H) = 7.57 dex.
This is in excellent agreement with our average O/H for the three
H\,{\sc ii} regions, which are roughly located in the same part of Sextans A.
Thus, the element abundances in these H\,{\sc ii} regions and in the three A
supergiants located in other parts of Sextans~A
\citep[see Figure~1 from][]{Kaufer04} do not provide any evidence
of metallicity inhomogeneity in this galaxy.

Since the abundances of the $\alpha$-elements S and Ar in the Sextans A
PN should not be affected by nucleosynthesis in its progenitor star,
these elements can be used as good tracers of the metallicity of the
interstellar gas in Sextans~A at the epoch when the PN progenitor was formed.
According to our estimate, this occurred some 1.6 Gyr ago 
(Table~\ref{tbl:PN_par}).
Due to the limited accuracy of S/H and Ar/H in the H\,{\sc ii} regions,
and especially
in the PN (in the latter the errors are on the order of about 40\% -- 50\%), the quantitative
result on the rate of the interstellar medium (ISM)
enrichment needs further improvement.
However, the values for relative enrichment, derived independently  for sulfur,
(S/H)$_{\rm HII}$/(S/H)$_{\rm PN}$ = 3.17 ($\pm ^{1.9}_{0.87}$)$\pm$1.05, and
argon,
(Ar/H)$_{\rm HII}$/(Ar/H)$_{\rm PN}$ = 2.99 ($\pm ^{2.83}_{0.98}$)$\pm$0.54,
are consistent with each other. We give the estimates of the uncertainties
of these ratios in the above form
since, due to the large relative errors of the values in the denominators,
1$\sigma$ intervals are highly non-symmetric; the values in parenthesis give
the uncertainties in the denominators, while the last term describes  
the uncertainties in the numerators.
The weighted average value is $\sim$3.1$\pm ^{2}_{1}$
or equivalently, 0.5$\pm$0.2 dex.

This rather large enrichment in $\alpha$-elements may be caused by
the increase in the star formation rate during the last $\sim$1--2 Gyr.
When we transform the above values to O/H and take the $\alpha$-element 
enrichment into account, we find 
12+$\log$(O/H)$_{\rm (-2~Gyr)}$ = 7.04$\pm$0.20 as the value prior to
enrichment.
The comparison of element abundances in three H\,{\sc ii} regions and in the PN
suggests that the region of Sextans~A where the PN progenitor
formed was as metal-deficient as I~Zw~18 is now \citep{SK93,IT98a,Kniazev03a}.
Moreover, assuming the element abundance pattern in the gas of which the PN
progenitor
formed to be typical of very low-metallicity H\,{\sc ii} galaxies,
we conclude that the
progenitor star, with an estimated mass of 1.5M$_\odot$, enriched the material
of this PN by a factor of $\sim$10 in
oxygen (from the value of 12+$\log$(O/H)=7.04 to the current value
of 8.02), and by a factor of $\sim$750 in nitrogen.

\subsubsection{Sextans B abundances}
\label{txt:disc_abun_SextB}

Previously, \citet{SCV96} observed, according to their description,
H\,{\sc ii}$-$5 (although the object indicated on their finding chart seems
to be closer to H\,{\sc ii}$-$10). They did not detect [O\,{\sc iii}]
$\lambda$4363 and presented a lower limit on O/H, equivalent to
12+$\log$(O/H)$\ge$7.4. Later, four H\,{\sc ii}
regions (H\,{\sc ii}$-$1, 2, 5, and 10) were observed by SKH89.
These authors failed to detect [O\,{\sc iii}] $\lambda$4363. To determine O/H,
they
used the empirical strong lines relation (which has an internal accuracy 
of 50\%) by \citet{Pagel89} and derived a lower limit for O/H in
H\,{\sc ii}$-$5.
Their resulting value for the oxygen abundance of Sextans B was 
12+$\log$(O/H)=7.56. Subsequently, MAM90 detected [O\,{\sc iii}] $\lambda$4363
in the spectrum of H\,{\sc ii}$-$5 at a level of 0.021$\pm$0.005
of I(H$\beta$)
and derived its O/H as 12+$\log$(O/H)=8.12, with no cited uncertainty.
We estimate that their rms is  $\ge0.1$ dex. It is worth noting here that the
empirical value of SKH89 for H\,{\sc ii}$-$5 is not in good agreement
with the value derived via the T$_{\rm e}$-method in MAM90.  In his
compilation, \citet{Mateo98} adopted an average value of
12+$\log$(O/H)=7.84$\pm$0.3 for Sextans~B.

\begin{figure}
    \begin{center}
    \epsscale{1.0}
    \includegraphics[clip=,angle=0,width=14cm]{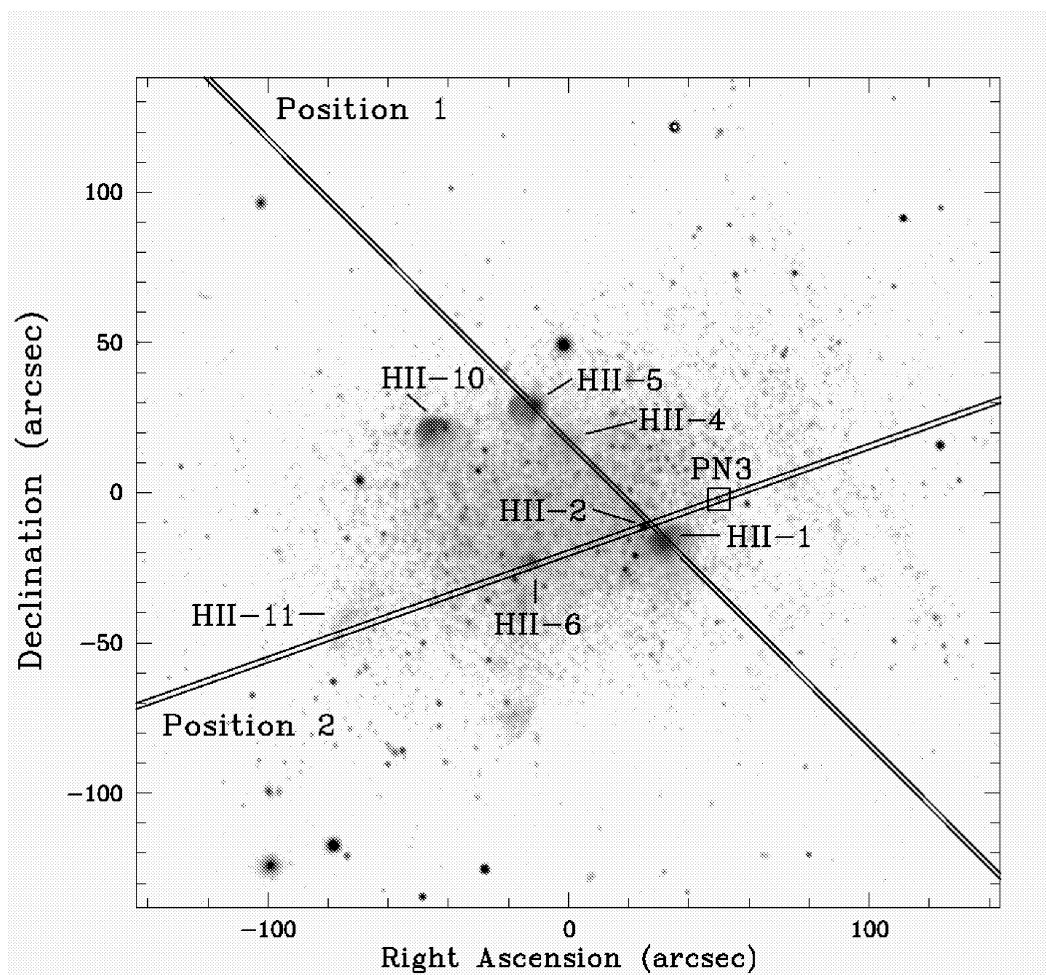}
    \caption{
     The NTT combined H$\alpha$ image of Sextans~B.
     North is up and East is to the left.
     The contrast is adjusted to highlight both bright and faint H$\alpha$
     emission.
     Black objects on the image indicate bright sources.
     Long-slit orientations are shown; the slit width is 2\arcsec.
     The position of PN3 is marked with a square. Labels indicate
     H\,{\sc ii} regions from \citet{SHK91} through which the slits were positioned, or which were mentioned in the text.
     At the adopted distance of 1.36 Mpc, 1\arcsec\ = 6.6 pc.
     The apparent line of emission in the upper right corner is
     an artifact resulting from CCD gaps in the mosaiced images.
    \label{fig:SexB_Ha}}
    \end{center}
\end{figure}

\begin{figure}
    \begin{center}
    \epsscale{1.0}
    \includegraphics[angle=0,width=18cm]{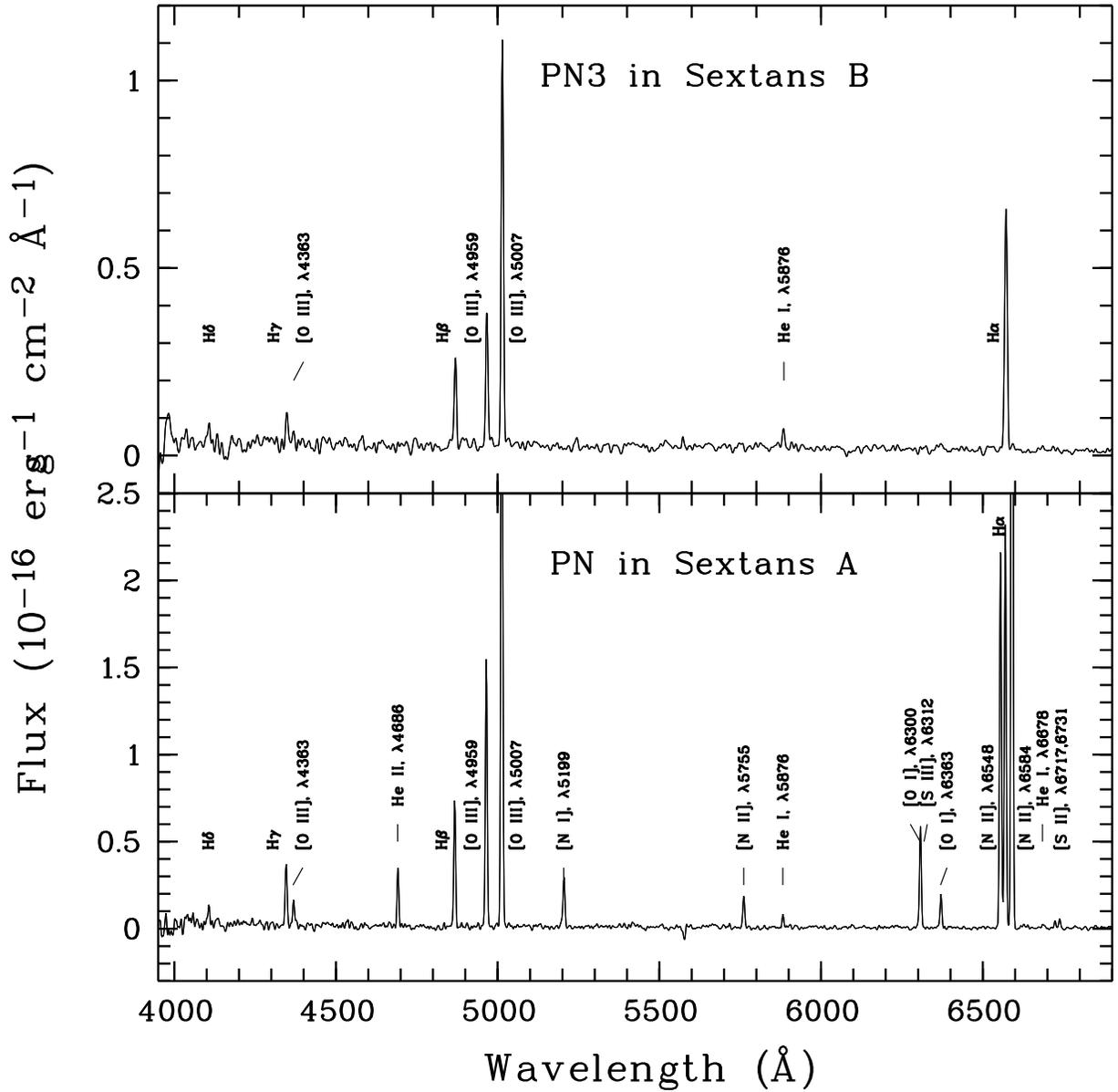}
    \caption{
    Emission-line spectra of the PN in Sextans~A and PN3 in Sextans~B
    from 4000 to 6900 \AA, obtained with grism \#5.
    The observed flux per unit wavelength is plotted versus wavelength.
    All detected emission lines are marked.
    \label{fig:SexA_PN_spec}}
    \end{center}
\end{figure}

\begin{figure}
    \begin{center}
    \epsscale{1.0}
    \includegraphics[angle=0,width=18cm]{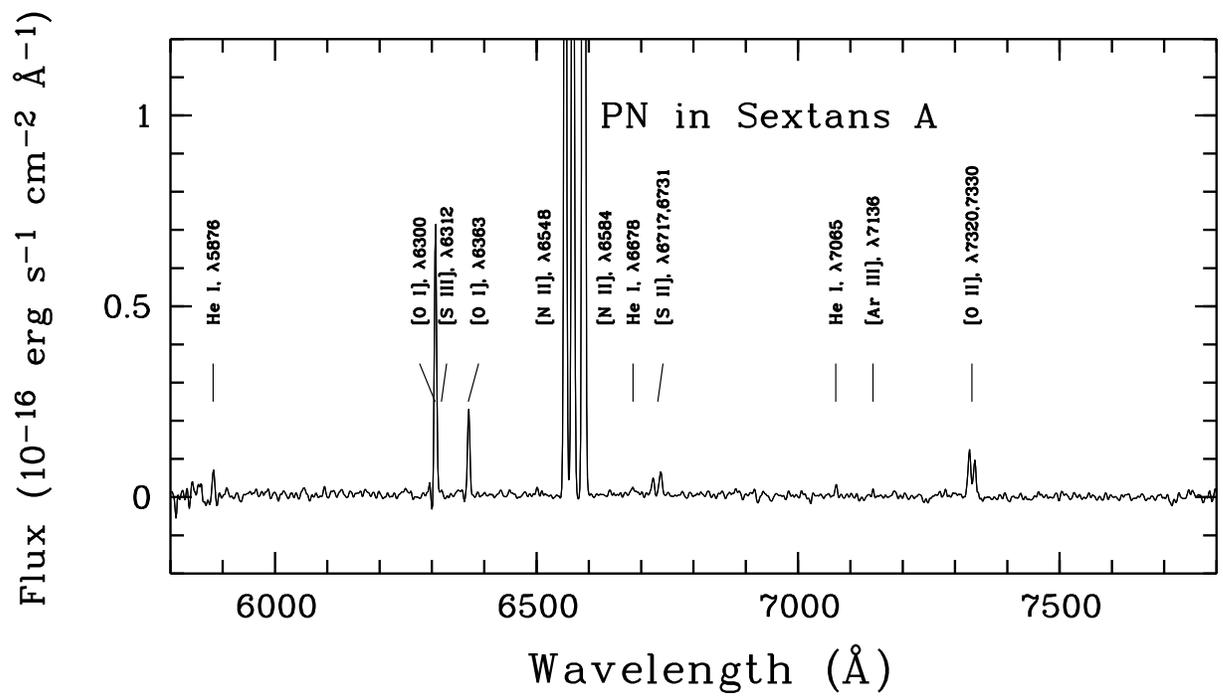}
    \caption{
     Emission-line spectrum of the PN in Sextans~A
     from 5800 to 7800 \AA, obtained with grism \#6.
    \label{fig:SexA_PN_spec_red}}
    \end{center}
\end{figure}

\begin{figure}
    \begin{center}
    \epsscale{1.0}
    \includegraphics[angle=0,width=14cm]{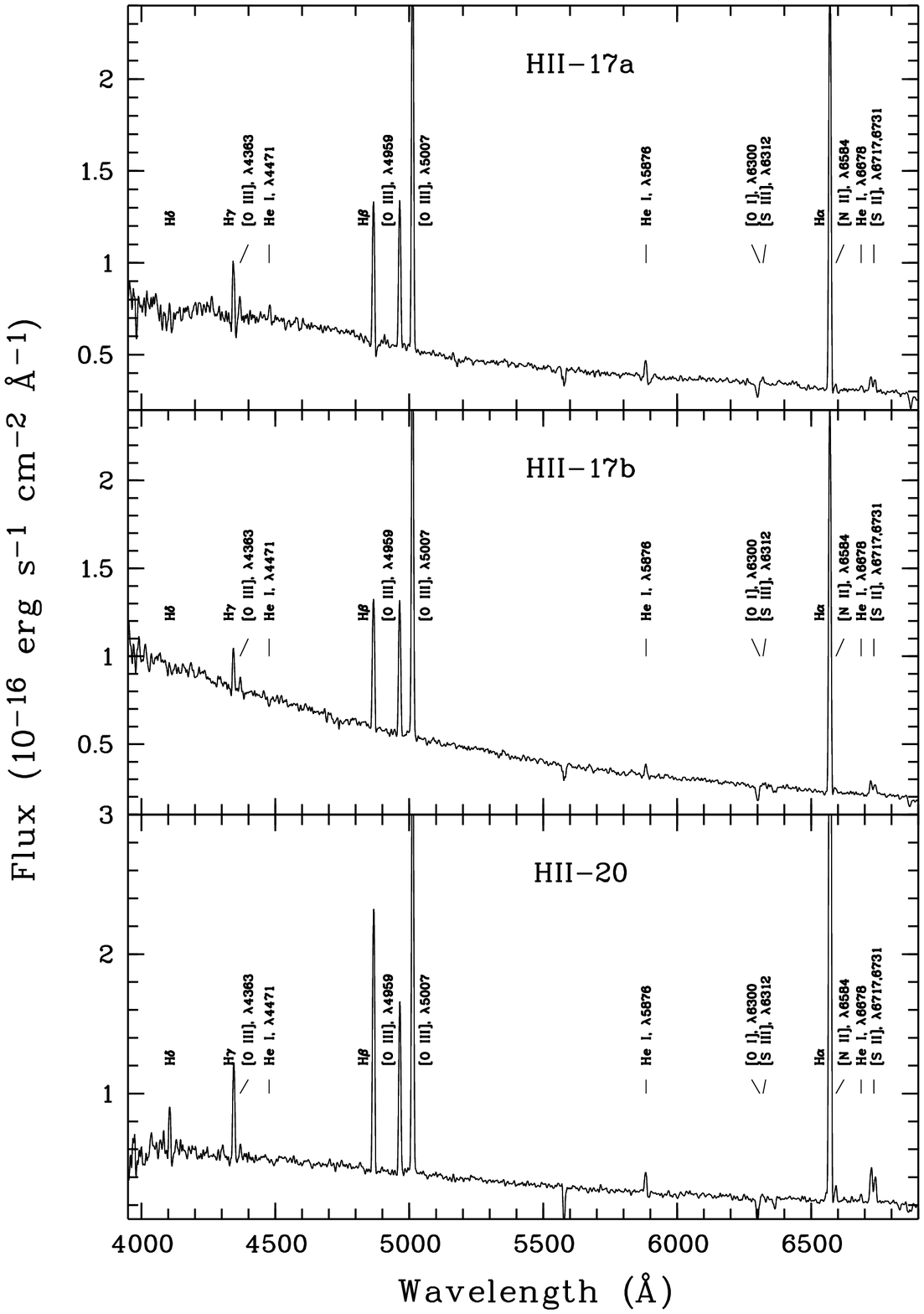}
    \caption{
    Emission-line spectra of H\,{\sc ii} regions in Sextans~A for which the
    [O~{\sc iii}] $\lambda$4363 line was detected,
    in the range from 4000 to 6900 \AA, obtained with grism \#5.
    \label{fig:SexA_HII_spec}}
    \end{center}
\end{figure}

\begin{figure}
    \begin{center}
    \epsscale{1.0}
    \includegraphics[angle=0,width=14cm]{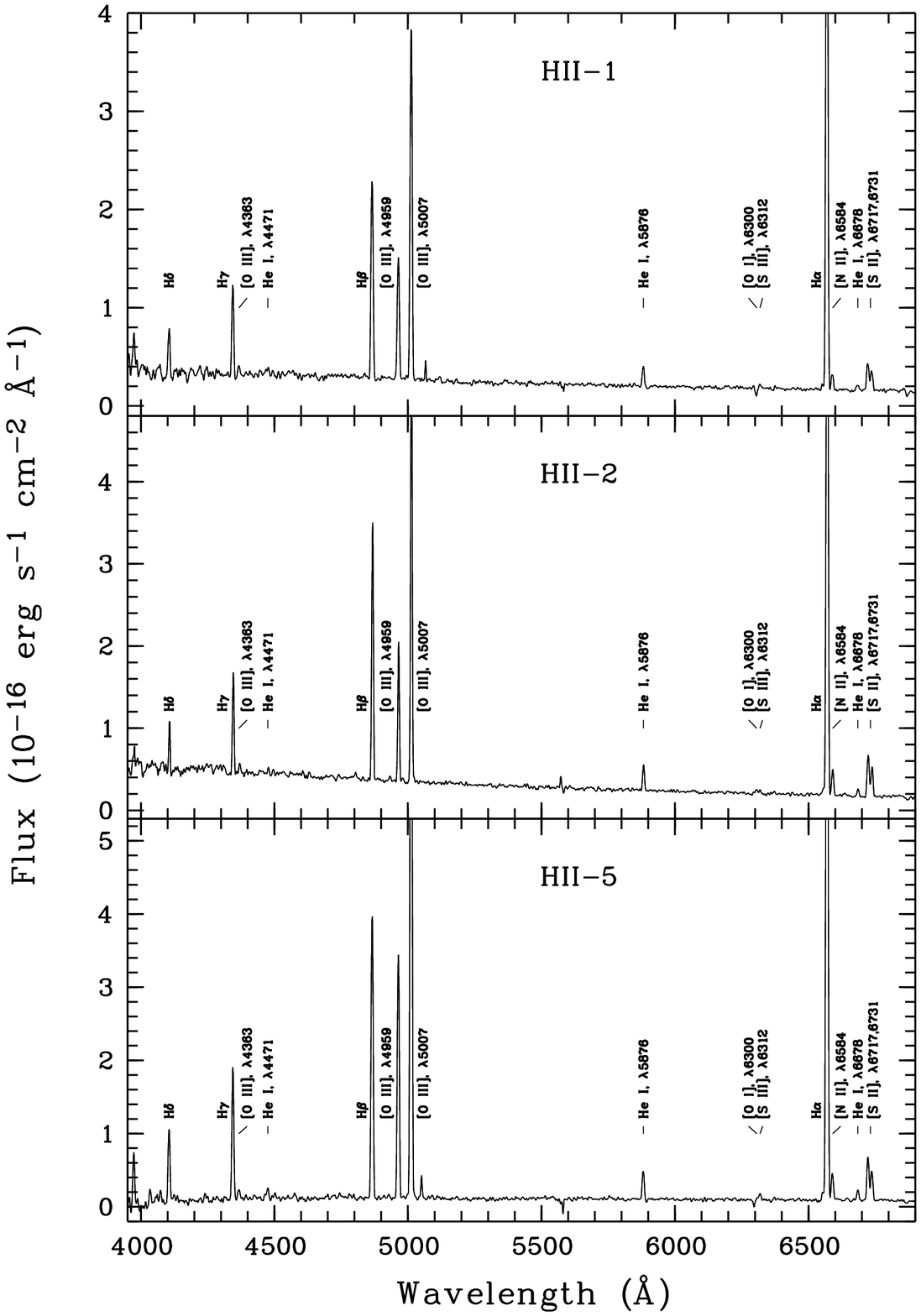}
    \caption{
    Emission-line spectra of H\,{\sc ii} regions in Sextans~B for which the
    [O~{\sc iii}] $\lambda$4363 line was detected, in the range
    from 4000 to 6900 \AA, obtained  with grism \#5.
    \label{fig:SexB_HII_spec}}
    \end{center}
\end{figure}

In our observations of Sextans~B H\,{\sc ii} regions,
the temperature-sensitive line [O\,{\sc iii}] $\lambda$4363 was detected,
and hence O/H was reliably measured, in only three of them.
For H\,{\sc ii}$-$1 and H\,{\sc ii}$-$2, which are
next to each other and which are 
situated at the western part of the brighter central
one-kpc region (Figure~\ref{fig:SexB_Ha}), the oxygen and other element
abundances are very similar.
For the third H\,{\sc ii} region (H\,{\sc ii}$-$5 in Figure \ref{fig:SexB_Ha},
at $\sim$0.6 kpc
NE) the oxygen abundance (12+$\log$(O/H)=7.84$\pm$0.05) is larger than for
the former two by a factor of 2.05$\pm$0.35. The ratios of the other elements
are similar:
(N/H)$_{\rm HII-5}$/(N/H)$_{\rm HII-1,2}=2.40\pm0.55$,
(Ar/H)$_{\rm HII-5}$/(Ar/H)$_{\rm HII-1,2}=2.89\pm0.50$, and
(S/H)$_{\rm HII-5}$/(S/H)$_{\rm HII-1,2}=3.33\pm0.8$.
All these ratios are consistent to within their rms uncertainties.  Their mean
abundances imply an enrichment by a factor of $\sim$2.5 relative to the element
abundances in H\,{\sc ii}$-$1 and 2. These differences are significant for each of the
four elements at a confidence level corresponding to 4~$\sigma_{\rm comb}$,
where $\sigma_{\rm comb}$ is the combined rms error of the difference
(on a linear scale) between the abundances in H\,{\sc ii}$-$5 and the average on
H\,{\sc ii}$-$1 and 2.

We have checked whether our results for H\,{\sc ii}$-$5 are consistent 
with the O/H
value from MAM90. For this we recalculated their O/H in our system,
using their relative
line intensities, and obtained 12+$\log$(O/H)=8.06. This is consistent
with our value to within $\sim$2$\sigma_{\rm comb}$. The main difference comes
from our higher (by a factor of 1.5) relative intensity of 
[O\,{\sc iii}] $\lambda$4363,
which is detected in our spectrum at an 8$\sigma$ level compared to a
4$\sigma$ detection in MAM90. Therefore, we consider our element
abundances to be more reliable. We also note that the apparent problem with
the small value of $\log$(N/O) = $-$1.8 in this H\,{\sc ii} region that
was pointed
out by \citet{SBK97} from the data of MAM90,
does not exist according to our data.

While the disagreement between the O/H values of SKH89 and
MAM90 could (in principle) be considered as an indication of
O/H inhomogeneity, our data provide the first {\it firm} evidence of
such chemical inhomogeneity in the interstellar medium of Sextans B
on scales of $\sim$0.5--1 kpc.

In contrast to the PN in Sextans~A,
the accuracy of the O/H and N/H ratios measured in PN3 in Sextans~B allows us
to formulate only very preliminary conclusions.
O/H and N/H in this PN are very close to the average values
derived for Sextans~B H\,{\sc ii} regions 1 and 2
(the difference is $\lesssim$0.5 $\sigma_{\rm comb}$).
The measured relative abundance log(N/O) = --1.48$\pm$0.24 for this PN
is well consistent with the average values for the most metal-poor
H\,{\sc ii} galaxies \citep{IT99}.
This suggests that the enrichment in O and N is sufficiently small
and in general does not exceed the uncertainty of log(N/O).
Adopting the 95\% confidence level for the
possible O and N abundances in PN3, we conclude that the maximum O and N
enrichment
since the epoch of the PN progenitor's formation does not exceed $\sim$0.4 dex.

\subsection{Chemical inhomogeneities in dwarf irregular galaxies}

Summarizing the results on the element abundances in Sextans B, we conclude
that our new spectroscopy confirms earlier conclusions (based on lower accuracy
measurements) that some of the nearby dIrr galaxies reveal considerable
abundance inhomogeneities among their H\,{\sc ii} regions.  This is an important
finding since dIrr galaxies are usually believed to be chemically well-mixed
and homogeneous at a given age.  To confidently determine the  
average ISM metallicity in a well-resolved dIrr galaxy one thus needs to 
measure its abundances in several regions.
Various processes (such as metal-enriched outflows, self-enrichment
on a time-scale of a few Myr, or infalling metal-poor intergalactic HI 
clouds) can in principle be responsible for the local deviations
of the ISM metallicity from some average galactic value. They may result
in either an excess or a deficiency as compared to the galactic mean. 
Note that in the lower-mass
dwarf spheroidal galaxies there are indications for chemical
inhomogeneities in terms of possible radial gradients among their
old stellar populations \citep{Harb01}. 
The available data on H\,{\sc ii} regions
of Sextans~B, taken at face value, are inconclusive in terms of
constraining the probable cause of the inhomogeneities.  It is unfortunate
that older tracers of the chemical evolutionary history of galaxies like
Sextans~B are so difficult to measure.  For the time being, only 
photometric estimates are available for the old population(s), based
either on isochrones or on globular cluster red giant branch fiducials.  
For Sextans~B it would be useful
to observe more H\,{\sc ii} regions at a sufficient depth to detect the
[O\,{\sc iii}] $\lambda$4363 line, in
order to explore what level of present-day 
metallicity is characteristic of the entire
galaxy. The other three H\,{\sc ii} regions visible in our
long-slit spectra will require observations with an 8-m class telescope to
reach an adequate signal to noise level.

It is worth noting that the census of H\,{\sc ii} region metallicities in dIrr
galaxies in the LG and its surroundings is rather
unsatisfactory. Only in the Magellanic Clouds can the number and spatial extent
of H\,{\sc ii} regions with measured element abundances be considered
sufficient for investigations of the chemical inhomogeneity problem
\citep[see, e.g.,][]{Pagel78,Vermeij03}. But even in the SMC, the
situation could be improved; while in the former study 19 H\,{\sc ii} regions
were studied across much of the radial extent of the Clouds, $\sim$5$^{\circ}$
($\sim$5 kpc), the uncertainties of the individual O/H values seem too
large to exclude possible variations at the level of $\sim$0.3 dex
(this level of variations we see in Sextans~B).
In the latter study the errors of O/H are reasonably small, and they are
consistent for all three measured H\,{\sc ii} regions to within 20\% (or 0.08 dex), but
all three regions are located close to each other
within a circle $\sim$20 pc in diameter.

For most of the 
other nearby dIrr galaxies, O/H values are based on old data,
generally of poor to moderate signal to noise for [O\,{\sc iii}] $\lambda$4363.
Alternatively (or in addition), in many cases only one to two 
H\,{\sc ii} regions per galaxy were measured
\citep[e.g.,][]{Pagel78,SKH89,STM89,MAM90,HM95,Lee03a,Lee03b}.
Thus the existing data for most nearby dIrr galaxies do not allow one
to draw a definite conclusion on chemical inhomogeneities in the ISM
at a level of $\sim$0.3 dex across the entire galaxy.
The general claim by \citet{Mateo98} that there is ``no evidence for
significant dispersion of oxygen abundances in any Local Group dIrrs in which
multiple H\,{\sc ii} regions have been studied'' should therefore be treated with some caution;
this statement probably applies only to variations with an amplitude of $\sim$0.4 dex
or higher.
We are aware of only a few similar occurrences in nearby dIrr galaxies:
in the LG dIrr galaxies SMC and  WLM and in the M~81 group dwarf Holmberg~II.
Star clusters in the SMC appear to differ in their metallicity at a
given age \citep{DaCosta02}, which seems to indicate differences in the
chemical composition in different regions of this galaxies at a given
time.  In WLM, \citet{Venn03} discovered that
one of two studied blue supergiants has a clearly higher [O/H] than the value
measured for the H\,{\sc ii} region WLM HM$-$7, situated 300 pc away from this
star.  The supergiant exceeds the [O/H] of the ISM by $\approx$0.7
dex at a 3$\sigma$ level, a curious difference for which an explanation still
needs to be found \citep{Lee05}.
For Holmberg~II we estimated from the data of \citet{Lee03b}
that the weighted average of 12+$\log$(O/H) for four H\,{\sc ii} regions
where [O\,{\sc iii}] $\lambda$4363 was measured is 7.63$\pm$0.08, while for the
H\,II$-$6 region the 2$\sigma$ lower limit is indicated as 7.97; in this galaxy several
H\,{\sc ii} regions show low abundances, while one appears to be enriched by a factor
of $\sim$2.

\citet{Roy95}, in their analysis of dispersal and mixing of oxygen
in the ISM of gas-rich galaxies, gave a fairly low characteristic time of 1.5
Myr for local mixing of freshly ejected nucleosynthesis products due to
Raleigh-Taylor and Kelvin-Helmholtz instabilities in star-forming regions.
While these processes are likely to be very efficient in mixing freshly
released processed heavy elements into the ambient ISM, detailed numerical
simulations would be desirable that take the specific conditions in our
target galaxies into account.
In their analysis of a large sample of H\,{\sc ii}
galaxies, \citet{SI03} also conclude that some self-enrichment should play a role in these
objects in order to match all observed correlations.

Galaxies of different types generally follow a trend of increased global
metallicity with increased luminosity, although offsets exist between 
different galaxy types \citep[e.g.,][GGH03 hereafter]{GGH03}.
The lower O/H in two H\,{\sc ii} regions of Sextans B
gives a better agreement with the empirical relation between O/H and
the blue luminosity for dwarf irregular galaxies (e.g., SKH89),
while the higher value of O/H in the 
H\,{\sc ii}$-$5 region shifts Sextans~B well above the 12+$\log$(O/H) vs
$L_{\rm B}$ relation.
Thus, the hypothesis of a significant overabundance in the region H\,{\sc ii}$-$5,
related to  localized metal pollution, sounds more realistic.
However, as discussed earlier, so far
such cases appear to be quite rare. This is in some aspects similar to the
detection of a significant nitrogen overabundance (on the spatial scale
of $\sim$50 pc) in the central starburst of a nearby dwarf starburst 
galaxy, NGC 5253 \citep{Kobul97}, again supporting very short
characteristic timescales of local enrichment and subsequent
dispersal as suggested by \citet{Roy95}. This may then imply
that we caught the evolution of H\,{\sc ii}$-$5 during the short period
when fresh $\alpha$-elements were just mixed into the $\sim$10$^{4}$~K medium
via stellar winds and supernova ejecta, but prior to their dilution 
via the mechanisms discussed, e.g., by \citet{Roy95}
and \citet{Tenorio96}.

If this is the case, the detection of chemical inhomogeneities in the
star-forming regions across the body of a dwarf galaxy is facilitated by the high
spatial resolution of our study ($\sim$6 pc), which avoids the effective
smearing of the localized short time-scale enrichment in more distant
actively star-forming galaxies due to limited angular/spatial resolution.
Obviously, this also implies that galaxies with many H\,{\sc ii} regions offer a
better chance to detect inhomogeneities than dIrrs with only one or two
H\,{\sc ii} regions.

In order to better understand possible reasons for the observed metal excess
in H\,{\sc ii}$-$5, more detailed and higher angular resolution studies of this region
would be helpful. If the excess is indeed related to the recent effective
mixing of the H\,{\sc ii} region's gas with matter from a hot bubble, some tracers of such
a process could be imprinted in the ionized gas kinematics, or in additional
ionization and excitation compared to more typical cases.

\subsection{Chemical evolution: Photometry versus spectroscopy}

\subsubsection{Sextans A}

How do the spectroscopic results compare with the 
photometric enrichment history derived by \citet{Dolphin03b}?
Using synthetic color-magnitude diagram (CMD) techniques, these authors
found a fairly constant average stellar metallicity throughout the galaxy's
history of [Fe/H] $\approx$ $-$1.45$\pm$0.2.  They suggest that most of the
enrichment occurred more than 10 Gyr ago. For the old red giant branch
GGH03 quote an approximate photometric mean metallicity of
$-1.9$ dex, estimated from comparison with globular cluster red giant
branch fiducials.  \citet{Dolphin03b}'s star formation history implies
that more than half of the stellar mass formed during that early period. 
However, \citet{Dolphin03b} also caution that the early star formation history
cannot be accurately inferred from the available HST data.
Approximately 10 Gyr ago the average star formation (SF hereafter) rate dropped
for a period of $\sim$7.5 Gyr by a factor
of $\sim$20 with only about 5\% of all stellar mass forming during that 
intermediate-age period.  Presumably, the low SF intensity
would have been accompanied by little chemical enrichment.  About 1--2
Gyr ago the SF again increased drastically, resulting in the formation
of about 30\% of the stellar mass of Sextans~A. This latter episode of
active star formation should have been
accompanied by a significant production of metals.  

The present-day spectroscopic metallicity of [Fe/H] $\approx$ $-$1.1 dex
(from H\,{\sc ii} regions and young A supergiants) is consistent
with an enrichment of about 0.4 dex
compared to the average stellar metallicity from \citet{Dolphin03b},
and with an enrichment of $\sim 0.8$ dex compared to the mean metallicity
of the old red giants in Sextans~A \citep{GGH03}.
One cannot exclude, however, that
due to the powerful energy release over the period of increased star
formation during the past 1--2 Gyr a significant fraction of freshly
synthesized heavy elements was blown
away. Thus, the metallicity in the young population and ISM may indicate
only a lower limit of the true metal production during this period.

In the above scenario, the results for the star formation history and
metal enrichment of Sextans~A, derived from photometric data, are compatible with our
conclusion of 0.5$\pm$0.2 dex enrichment according to the comparison of the
H\,{\sc ii} region abundances and the data on S and Ar abundances
in the sole PN of Sextans~A.
As we show in Table~\ref{tbl:PN_par} the PN progenitor was a low-mass star with
$\sim$1.5~M$_{\odot}$ and an age of $\sim$1.6 Gyr, i.e., it
formed at that time when the star formation rate and the
accompanying enrichment showed a marked increase, according to
\citet{Dolphin03b}.

\subsubsection{Sextans B}

For Sextans B, the deepest published CMD is based on HST data \citep{Kar02}
but only covers the upper red giant branch.  Hence no attempt was made to
derive the star formation history of Sextans B from these data.  However,
the CMD of \citet{Kar02} allows us to draw the following qualitative
conclusions:  Sextans B possesses a substantial intermediate-age to old
population with ages exceeding $\sim 2$ Gyr as evidenced by its very well 
populated, prominent red giant branch.
Our estimate of the age of the observed PN progenitor (see Table~\ref{tbl:PN_par})
also supports the idea that Sextans B had increased SF about 6 Gyr ago.
GGH03 suggest a mean
photometric metallicity of $-2.1$ dex for the old population
through comparison of the red giant branch with globular cluster fiducials.
The CMD of \citet{Kar02} also shows the presence of a large number of 
luminous asymptotic giant branch stars above the tip of the red giant
branch, and an extended group of young blue main-sequence stars and 
blue-loop stars.  Red supergiants appear to be present as well.  To
summarize, the CMD shows the typical features also known from many other
dIrr galaxies, which support the scenario of star formation continuing
over a Hubble time (e.g., Grebel 1999).  In earlier work based on 
much shallower ground-based
data, \citet{Tos91} studied the recent star formation history of Sextans B
up to 1 Gyr and found moderate
star formation activity, probably intermittent.
They provide a rough photometric estimate of the young population's stellar 
metallicity of $-1.3$ dex.  While they claim
that the oxygen abundance in the H\,{\sc ii}$-$10 region of
12+$\log$(O/H)=8.1 (MAM90) is in the perfect agreement
with their stellar metallicity estimate, this is no longer correct. 
In the modern calibration the above oxygen abundance
corresponds to [O/H]=$-$0.65.
The current ISM metallicity, as measured here by O/H in
the two low-metallicity H\,{\sc ii} regions, corresponds to
[O/H] = $-$1.13$\pm$0.07.
The latter, however, is in agreement with the estimate of the young 
population's stellar metallicity from \citet{Tos91}. We estimated 
in Section~\ref{txt:disc_abun_SextB}
that the maximum O and N enrichment during the last few
Gyr does not exceed $\sim$0.4 dex for Sextans~B.
If this is true, one can suggest that most
of the enrichment occured in Sextans~B
more than 6 Gyr ago, as would also be suggested by the very low
metallicity of the old red giant branch in Sextans~B derived by
GGH03.

The qualitative star formation history 
of Sextans B over the total cosmological time, based on all previously
available CMD data is summarized by \citet{Gre97}, \citet{Mateo98}, and
\citet{Gre99}. It may have shown a pronounced early star formation rate
during the first few Gyr with a
subsequent period of possibly lower star formation activity 
and a potential increase at more recent times (say, 1--2 Gyr ago).
It should be emphasized that there is still very little quantitative
information on the star formation history of Sextans~B.

\subsection{Sextans~A versus Sextans~B SF histories}

The general similarity of the SF histories of both Sextans A and B and 
other nearby dwarf irregulars is important in the aspect of dwarf galaxy
evolution in small galaxy 
groups. This likely implies that the SF drivers are similar
in the
LG and this particular dwarf group. Another question concerns the
nature of significant differences in SF histories during the last several Gyr; since there appears to be little correlation between the SF histories of Sextans~A and B, it is tempting
to suggest that the
SF during this period was mainly due to intrinsic processes, unrelated to the
tidal action
from a nearby neighbor. Indeed, the estimate of tidal forces between the two
dwarfs, based on their projected distance of $\sim$250 pc, total masses
of $\sim$(4--8)$\times$10$^{8}$ M$_{\odot}$ and Holmberg radii of $\sim$0.8--0.9
kpc \citep{Mateo98} results in a relative strength of tides of the order of
10$^{-7}$.  However, this issue needs more careful study including the
morphology and kinematics of gas and stars in these galaxies, and more
importantly, constraints on their orbits.

While the SF histories of dwarf galaxies of the same morphological type 
exhibit similar overall properties and trends, they differ in the details.
Indeed, each dwarf has its own unique evolutionary history (e.g., Grebel
1997).  These differences become most pronounced on scales of a fraction
of a Gyr, but may encompass time scales of several Gyr.  In the absence
of detailed, deep CMDs and detailed information on the chemical evolution
of such galaxies, the PN census may serve as a possible indicator.
Since the absolute magnitudes $M_{\rm B}$ of Sextans A and B are very similar
(see Table~\ref{tbl:General}), and since their global SF histories appear
to be similar, we should then expect that both galaxies also have a 
comparable number of PNe.  Instead, surveys for PNe in the two dIrrs have
revealed a difference by a factor of 5 in the PN census.  This may imply 
that Sextans~B experienced a significantly higher SF rate during the 
aforementioned time interval. Similar conclusions the likely 
differing SF histories in Sextans~A and B during last few Gyr were reached
by \citet{Magr02,Magr03,Magr04}.

\section{Summary and conclusions}
\label{txt:summ}

In this paper we presented new determinations of the chemical
abundances of six H\,{\sc ii} regions in Sextans A and B, and of one PN in each
of these galaxies.
Based on the data and discussion presented in the paper, the following
conclusions can be drawn:

     1. We confirmed with good accuracy the low ISM metallicity of both
     galaxies, with 12+$\log$(O/H) = 7.54$\pm$0.06 in three H\,{\sc ii} regions in
     Sextans A, and 12+$\log$(O/H) = 7.53$\pm$0.05 in two H\,{\sc ii} regions in
     Sextans B. The element abundance ratios of O, N, S, and Ar are well
     consistent with the expected
     patterns of very metal-poor
     H\,{\sc ii} galaxies.

     2. New high accuracy chemical abundances in the H\,{\sc ii} regions of
     Sextans A and B allowed us to probe the present-day
     metallicity homogeneity across the bodies of 
     the respective galaxies.
     While our statistics are still poor
     (three H\,{\sc ii} regions per
     galaxy), we find that in Sextans A the measured abundances in all
     three H\,{\sc ii} regions show no differences exceeding 0.1 dex (at the 63\% confidence level).
     Moreover, the metallicities of three A-supergiants studied 
     by \citet{Kaufer04} are in good agreement with those of the H\,{\sc ii} regions.

     3. In Sextans~B one H\,{\sc ii} region (H\,{\sc ii}$-$5) is significantly enriched, with
     an  excess in O, N, S, and Ar of a factor of 2.5$\pm$0.5
     relative to the mean value of the two other H\,{\sc ii} regions studied
     here.  This is strong evidence for chemical inhomogeneity in a dIrr
     galaxy.  Whether this implies a general chemical
     inhomogeneity among populations of comparable age in Sextans B, and thus
     a metallicity spread at a given age, or whether we happen to see the
     short-lived effects of freshly ejected nucleosynthesis products prior
     to their dispersal and mixing with the ambient interstellar medium
     will require further study.

    4. Chemical abundances derived for the PN in Sextans~A show that this
    is a Type~I object with a highly elevated nitrogen abundance: N/O $\sim$2.5.
    Its O/H is about a factor of 3 higher than in the H\,{\sc ii} regions, 
    which implies
    significant self-pollution by the PN progenitor.
    The abundances of S and Ar indicate that the ISM metallicity
    was $\sim$0.5 dex lower at the time of the formation of the PN progenitor,
    compared to that currently measured in H\,{\sc ii} regions.
    In this case the PN progenitor enriched the material by a factor of $\sim$10
    in oxygen, and by a factor of $\sim$750 in nitrogen.

    5. The element abundances of PN3 in Sextans B are consistent with
    those of the two H\,{\sc ii} regions with low metallicity.
    Comparison of the element abundances for this PN
    and the two nearby H\,{\sc ii} regions implies that the maximum
    O and N enrichment during the last few 
    Gyr (the estimated age of the progenitor star of PN3) did not exceed
    $\sim$0.4 dex.

    6. Despite the overall similarity of Sextans A and B, their star
    formation history and enrichment history over the last few Gyr 
    looks quite different, as evidenced by both the
    number of H\,{\sc ii} regions and PNe, and their abundances.
    The overall chemical enrichment experienced by Sextans A and B
    as judged from the estimated mean metallicity of their old red giant
    branch populations and their H\,{\sc ii} regions has spanned at least
    0.8 dex.  This number is a lower limit since it refers to the mean
    red giant metallicity and necessarily neglects the still unknown
    metallicity of the most metal-poor giants in the two dIrrs.

\acknowledgments

The authors are pleased to thank L.A. Pustilnik for
consultations on plasma instabilities. We thank also the anonymous referee
for useful comments and suggestions.
S.A.P. and A.G.P. acknowledge financial support and hospitality of MPIA 
during part of this work.
EKG was supported by the Swiss National Science Foundation through
the grants 200021-101924/1 and 200020-105260/1.

This research has made use of the
NASA/IPAC Extragalactic Database (NED) which is operated by the Jet
Propulsion Laboratory, California Institute of Technology, under contract
with the National Aeronautics and Space Administration.

\clearpage
%
%

\clearpage
%
%

\begin{deluxetable}{lcc}
\tablecaption{Basic Parameters of the Sextans A and Sextans B Dwarf Irregular Galaxies
	      \label{tbl:General}}
\tablewidth{0pt}
\tablehead{
\colhead{Property} & \colhead{Sextans A} & \colhead{Sextans B} \\
\colhead{(1)}      & \colhead{(2)}       & \colhead{(3)}
}
\startdata
Alternate Names              & UGCA~205, DDO~75                              & UGC~5373, DDO~70                             \\
$\alpha$ (J2000.0)           & 10$^h$11$^m$00.80$^s$                         & 10$^h$00$^m$00.10$^s$                        \\
$\delta$ (J2000.0)           & $-$04$^\circ$41$^\prime$34.0$^{\prime\prime}$ & +05$^\circ$19$^\prime$56.0$^{\prime\prime}$  \\
Distance (kpc)               & 1320$\pm$40$^1$                               & 1360$\pm$70$^2$                              \\
V$_\odot$,opt (km s$^-{1}$)  & 328$\pm$5$^3$                                 & 301$\pm$1$^5$                                \\
V$_\odot$,radio (km s$^-{1}$)& 325$\pm$3$^4$                                 & 303$\pm$2$^6$                                \\
B$_T$ (mag)                  & 11.86$^7$                                     & 11.85$^2$                                    \\
A$_B$ (mag)$^a$              & 0.19                                          & 0.14                                         \\
(m-M$_0$) (mag)              & 25.62$^7$                                     & 25.67$^7$                                    \\
M$_B$                        & $-13.96$~                                     & $-13.97$~                                    \\
\enddata
\tablerefs{(1) \citet{Dolphin03a};
	   (2) \citet{Kar02};
	   (3) \citet{TOS93};
	   (4) \citet{STTW88};
	   (5) \citet{Fal99};
	   (6) \citet{Hof96};
	   (7) \citet{Kar04}
}
\tablenotetext{a}{Galactic extinction in the $B$-band is from NED, based on \citet{Schlegel98}.}
\end{deluxetable}

%
%
\begin{deluxetable}{lcccccc}
\tabletypesize{\footnotesize}
\tablecaption{Line intensities of H\,{\sc ii} regions in Sextans A
	      \label{tbl:SextA_HII_int}}
\tablewidth{0pt}
\tablehead{
& \MC{2}{c}{H\,{\sc ii}$-$20}  &  \MC{2}{c}{H\,{\sc ii}$-$17a}  &  \MC{2}{c}{H\,{\sc ii}$-$17b}  \\ \hline
$\lambda_{0}$(\AA) Ion                    & F($\lambda$)/F(H$\beta$)&I($\lambda$)/I(H$\beta$)
					  & F($\lambda$)/F(H$\beta$)&I($\lambda$)/I(H$\beta$)
					  & F($\lambda$)/F(H$\beta$)&I($\lambda$)/I(H$\beta$) \\
\colhead{(1)} & \colhead{(2)} & \colhead{(3)} & \colhead{(4)} & \colhead{(5)} & \colhead{(6)} & \colhead{(7)}
}
\startdata
4101\ H$\delta$\                          & 0.191$\pm$0.009 & 0.285$\pm$0.019 & \nodata         & \nodata         & \nodata         & \nodata         \\
4340\ H$\gamma$\                          & 0.381$\pm$0.011 & 0.458$\pm$0.017 & 0.404$\pm$0.020 & 0.472$\pm$0.031 & 0.391$\pm$0.030 & 0.470$\pm$0.045 \\
4363\ [O\ {\sc iii}]\                     & 0.034$\pm$0.006 & 0.034$\pm$0.007 & 0.090$\pm$0.015 & 0.085$\pm$0.016 & 0.092$\pm$0.021 & 0.087$\pm$0.021 \\
4861\ H$\beta$\                           & 1.000$\pm$0.018 & 1.000$\pm$0.020 & 1.000$\pm$0.032 & 1.000$\pm$0.038 & 1.000$\pm$0.039 & 1.000$\pm$0.045 \\
4959\ [O\ {\sc iii}]\                     & 0.680$\pm$0.015 & 0.626$\pm$0.015 & 1.025$\pm$0.032 & 0.923$\pm$0.032 & 1.009$\pm$0.040 & 0.930$\pm$0.040 \\
5007\ [O\ {\sc iii}]\                     & 1.849$\pm$0.027 & 1.695$\pm$0.027 & 2.916$\pm$0.070 & 2.616$\pm$0.069 & 3.030$\pm$0.090 & 2.801$\pm$0.090 \\
5876\ He\ {\sc i}\                        & 0.077$\pm$0.009 & 0.065$\pm$0.008 & 0.178$\pm$0.017 & 0.151$\pm$0.016 & 0.105$\pm$0.024 & 0.094$\pm$0.022 \\
6300\ [O\ {\sc i}]\                       & 0.027$\pm$0.019 & 0.022$\pm$0.017 & \nodata         & \nodata         & \nodata         & \nodata         \\
6312\ [S\ {\sc iii}]\                     & 0.008$\pm$0.014 & 0.007$\pm$0.013 & 0.035$\pm$0.013 & 0.029$\pm$0.011 & \nodata         & \nodata         \\
6548\ [N\ {\sc ii}]\                      & 0.029$\pm$0.007 & 0.024$\pm$0.006 & \nodata         & \nodata         & \nodata         & \nodata         \\
6563\ H$\alpha$\                          & 3.444$\pm$0.045 & 2.803$\pm$0.043 & 3.288$\pm$0.077 & 2.746$\pm$0.078 & 3.009$\pm$0.087 & 2.678$\pm$0.091 \\
6584\ [N\ {\sc ii}]\                      & 0.065$\pm$0.011 & 0.052$\pm$0.010 & 0.084$\pm$0.018 & 0.068$\pm$0.017 & 0.055$\pm$0.024 & 0.048$\pm$0.022 \\
6678\ He\ {\sc i}\                        & 0.031$\pm$0.008 & 0.025$\pm$0.007 & 0.064$\pm$0.022 & 0.052$\pm$0.020 & \nodata         & \nodata         \\
6717\ [S\ {\sc ii}]\                      & 0.150$\pm$0.012 & 0.119$\pm$0.010 & 0.135$\pm$0.020 & 0.110$\pm$0.018 & 0.110$\pm$0.024 & 0.096$\pm$0.023 \\
6731\ [S\ {\sc ii}]\                      & 0.104$\pm$0.010 & 0.083$\pm$0.009 & 0.098$\pm$0.014 & 0.079$\pm$0.013 & 0.098$\pm$0.021 & 0.086$\pm$0.020 \\
7065\ He\ {\sc i}\                        & 0.026$\pm$0.010 & 0.020$\pm$0.009 & \nodata         & \nodata         & \nodata         & \nodata         \\
7136\ [Ar\ {\sc iii}]\                    & 0.069$\pm$0.013 & 0.053$\pm$0.011 & 0.084$\pm$0.022 & 0.067$\pm$0.020 & 0.088$\pm$0.016 & 0.076$\pm$0.015 \\
7325\ [O\ {\sc ii}]\                      & 0.047$\pm$0.010 & 0.036$\pm$0.009 & 0.096$\pm$0.022 & 0.076$\pm$0.020 & 0.090$\pm$0.021 & 0.078$\pm$0.020 \\
 & & & & & & \\
C(H$\beta$)\ dex              & \MC {2}{c}{0.19$\pm$0.02} & \MC {2}{c}{0.14$\pm$0.03} & \MC {2}{c}{0.07$\pm$0.04} \\
EW(abs)\ \AA\                 & \MC {2}{c}{3.3$\pm$0.3}   & \MC {2}{c}{1.7$\pm$0.3}   & \MC {2}{c}{1.1$\pm$0.3}   \\
EW(H$\beta$)\ \AA\            & \MC {2}{c}{  43$\pm$ 1}   & \MC {2}{c}{16$\pm$1}      & \MC {2}{c}{13$\pm$1}      \\
F(H$\beta$)\tablenotemark{a}\ & \MC {2}{c}{18.4$\pm$0.2}  & \MC {2}{c}{7.2$\pm$0.2}   & \MC {2}{c}{7.7$\pm$0.2}   \\ \hline
\enddata
\tablenotetext{a}{Observed flux in units of 10$^{-16}$ ergs\ s$^{-1}$cm$^{-2}$}
\end{deluxetable}

%
%
\begin{deluxetable}{lcccccc}
\tabletypesize{\footnotesize}
\tablecaption{Line intensities of H\,{\sc ii} regions in Sextans B
	      \label{tbl:SextB_HII_int}}
\tablewidth{0pt}
\tablehead{
& \MC{2}{c}{H\,{\sc ii}$-$5}  &  \MC{2}{c}{H\,{\sc ii}$-$1}  &  \MC{2}{c}{H\,{\sc ii}$-$2}  \\ \hline
$\lambda_{0}$(\AA) Ion                    & F($\lambda$)/F(H$\beta$)&I($\lambda$)/I(H$\beta$)
					  & F($\lambda$)/F(H$\beta$)&I($\lambda$)/I(H$\beta$)
					  & F($\lambda$)/F(H$\beta$)&I($\lambda$)/I(H$\beta$) \\
\colhead{(1)} & \colhead{(2)} & \colhead{(3)} & \colhead{(4)} & \colhead{(5)} & \colhead{(6)} & \colhead{(7)}
}
\startdata
3868\ [Ne\ {\sc iii}]\                    & 0.136$\pm$0.005 & 0.137$\pm$0.005  & \nodata         & \nodata         & \nodata         & \nodata          \\
3889\ He\ {\sc i}\ +\ H8\                 & 0.203$\pm$0.006 & 0.219$\pm$0.008  & 0.184$\pm$0.009 & 0.241$\pm$0.014 & \nodata         & \nodata          \\
3967\ [Ne\ {\sc iii}]\ +\ H7\             & 0.179$\pm$0.005 & 0.182$\pm$0.005  & \nodata         & \nodata         & \nodata         & \nodata          \\
4101\ H$\delta$\                          & 0.230$\pm$0.004 & 0.244$\pm$0.006  & 0.205$\pm$0.007 & 0.264$\pm$0.012 & 0.164$\pm$0.006 & 0.274$\pm$0.013  \\
4340\ H$\gamma$\                          & 0.461$\pm$0.008 & 0.473$\pm$0.009  & 0.430$\pm$0.015 & 0.477$\pm$0.019 & 0.367$\pm$0.008 & 0.460$\pm$0.012  \\
4363\ [O\ {\sc iii}]\                     & 0.033$\pm$0.004 & 0.032$\pm$0.004  & 0.045$\pm$0.008 & 0.043$\pm$0.008 & 0.040$\pm$0.005 & 0.038$\pm$0.005  \\
4471\ He\ {\sc i}\                        & 0.038$\pm$0.004 & 0.037$\pm$0.004  & \nodata         & \nodata         & 0.023$\pm$0.004 & 0.022$\pm$0.004  \\
4740\ [Ar\ {\sc iv]}\                     & 0.018$\pm$0.003 & 0.017$\pm$0.003  & \nodata         & \nodata         & \nodata         & \nodata          \\
4861\ H$\beta$\                           & 1.000$\pm$0.009 & 1.000$\pm$0.010  & 1.000$\pm$0.013 & 1.000$\pm$0.015 & 1.000$\pm$0.013 & 1.000$\pm$0.015  \\
4959\ [O\ {\sc iii}]\                     & 0.843$\pm$0.009 & 0.827$\pm$0.009  & 0.616$\pm$0.012 & 0.579$\pm$0.012 & 0.530$\pm$0.009 & 0.483$\pm$0.009  \\
5007\ [O\ {\sc iii}]\                     & 2.524$\pm$0.017 & 2.474$\pm$0.017  & 1.752$\pm$0.019 & 1.644$\pm$0.019 & 1.613$\pm$0.018 & 1.463$\pm$0.018  \\
5876\ He\ {\sc i}\                        & 0.107$\pm$0.004 & 0.103$\pm$0.004  & 0.104$\pm$0.009 & 0.096$\pm$0.009 & 0.108$\pm$0.007 & 0.091$\pm$0.006  \\
6300\ [O\ {\sc i}]\                       & \nodata         & \nodata          & \nodata         & \nodata         & 0.020$\pm$0.007 & 0.017$\pm$0.006  \\
6312\ [S\ {\sc iii}]\                     & 0.023$\pm$0.002 & 0.022$\pm$0.002  & 0.014$\pm$0.003 & 0.013$\pm$0.003 & 0.014$\pm$0.004 & 0.012$\pm$0.004  \\
6548\ [N\ {\sc ii}]\                      & 0.028$\pm$0.003 & 0.027$\pm$0.003  & 0.028$\pm$0.006 & 0.026$\pm$0.005 & 0.037$\pm$0.005 & 0.029$\pm$0.004  \\
6563\ H$\alpha$\                          & 2.952$\pm$0.028 & 2.829$\pm$0.030  & 3.003$\pm$0.056 & 2.766$\pm$0.060 & 3.387$\pm$0.041 & 2.770$\pm$0.040  \\
6584\ [N\ {\sc ii}]\                      & 0.097$\pm$0.004 & 0.092$\pm$0.004  & 0.084$\pm$0.008 & 0.076$\pm$0.007 & 0.111$\pm$0.009 & 0.090$\pm$0.008  \\
6678\ He\ {\sc i}\                        & 0.036$\pm$0.003 & 0.034$\pm$0.002  & 0.036$\pm$0.008 & 0.033$\pm$0.007 & 0.037$\pm$0.006 & 0.030$\pm$0.005  \\
6717\ [S\ {\sc ii}]\                      & 0.168$\pm$0.004 & 0.160$\pm$0.004  & 0.154$\pm$0.008 & 0.140$\pm$0.008 & 0.206$\pm$0.007 & 0.165$\pm$0.006  \\
6731\ [S\ {\sc ii}]\                      & 0.115$\pm$0.004 & 0.110$\pm$0.004  & 0.108$\pm$0.008 & 0.098$\pm$0.007 & 0.146$\pm$0.006 & 0.117$\pm$0.006  \\
7065\ He\ {\sc i}\                        & 0.025$\pm$0.003 & 0.024$\pm$0.003  & 0.025$\pm$0.006 & 0.023$\pm$0.005 & 0.026$\pm$0.006 & 0.020$\pm$0.005  \\
7136\ [Ar\ {\sc iii}]\                    & 0.077$\pm$0.003 & 0.073$\pm$0.003  & 0.062$\pm$0.006 & 0.056$\pm$0.006 & 0.071$\pm$0.007 & 0.056$\pm$0.006  \\
7325\ [O\ {\sc ii}]\                      & 0.044$\pm$0.004 & 0.042$\pm$0.004  & 0.064$\pm$0.010 & 0.057$\pm$0.010 & 0.099$\pm$0.011 & 0.077$\pm$0.009  \\
 & & & & & & \\
C(H$\beta$)\ dex              & \MC {2}{c}{0.04$\pm$0.01} & \MC {2}{c}{0.04$\pm$0.02} & \MC {2}{c}{0.17$\pm$0.02} \\
EW(abs)\ \AA\                 & \MC {2}{c}{5.5$\pm$1.6}   & \MC {2}{c}{4.4$\pm$0.4}   & \MC {2}{c}{5.8$\pm$0.3}   \\
EW(H$\beta$)\ \AA\            & \MC {2}{c}{60$\pm$1}      & \MC {2}{c}{70$\pm$1}      & \MC {2}{c}{66$\pm$1}      \\
F(H$\beta$)\tablenotemark{a}\ & \MC {2}{c}{36.3$\pm$0.3}  & \MC {2}{c}{19.8$\pm$0.2}  & \MC {2}{c}{24.1$\pm$0.2 } \\ \hline
\enddata
\tablenotetext{a}{Observed flux in units of 10$^{-16}$ ergs\ s$^{-1}$cm$^{-2}$}
\end{deluxetable}

%
%
\begin{deluxetable}{lccc}
\tabletypesize{\footnotesize}
\tablecaption{Abundances in Sextans A H\,{\sc ii} regions
	      \label{tbl:SextA_HII_ab}}
\tablewidth{0pt}
\tablehead{
\colhead{Value}    & \colhead{H\,{\sc ii}$-$20} & \colhead{H\,{\sc ii}$-$17a} & \colhead{H\,{\sc ii}$-$17b} \\
\colhead{(1)}      & \colhead{(2)}   & \colhead{(3)}   & \colhead{(4)}
}
\startdata
$T_{\rm e}$(OIII)(K)\                & 15,000$\pm$1400~~  &  19,460$\pm$2180~~ & 19,040$\pm$2750~~    \\
$T_{\rm e}$(OII)(K)\                 & 13,760$\pm$1230~~  &  15,520$\pm$1620~~ & 15,390$\pm$2070~~    \\
$T_{\rm e}$(SIII)(K)\                & 14,160$\pm$1170~~  &  17,860$\pm$1800~~ & 17,500$\pm$2280~~    \\
$N_{\rm e}$(SII)(cm$^{-3}$)\         &  10$\pm$10 ~~      &   35$\pm$25~~      & 390$\pm$90~~         \\
& \\
O$^{+}$/H$^{+}$($\times$10$^5$)\     & 1.59$\pm$0.55~~    &  1.858$\pm$0.684~~ & 1.527$\pm$0.650~~    \\
O$^{++}$/H$^{+}$($\times$10$^5$)\    & 1.89$\pm$0.46~~    &  1.609$\pm$0.414~~ & 1.778$\pm$0.592~~    \\
O/H($\times$10$^5$)\                 & 3.48$\pm$0.72~~    &  3.467$\pm$0.800~~ & 3.305$\pm$0.879~~    \\
12+log(O/H)\                         & ~7.54$\pm$0.09~~   &  ~7.54$\pm$0.10~~  & ~7.52$\pm$0.11~~     \\
& \\
N$^{+}$/H$^{+}$($\times$10$^7$)\     & 4.56$\pm$1.01~~    &  4.75$\pm$1.23~~   & 3.43$\pm$1.46~~      \\
ICF(N)\                              & 2.19               &  1.87              & 2.16                 \\
N/H($\times$10$^7$)\                 & 9.98$\pm$2.22~~    &  8.87$\pm$2.29~~   & 7.43$\pm$3.16~~      \\
12+log(N/H)\                         & 6.00$\pm$0.10~~    &  5.95$\pm$0.11~~   & 5.87$\pm$0.18~~      \\
log(N/O)\                            & --1.54$\pm$0.13~~  &  --1.59$\pm$0.15~~ & --1.65$\pm$0.21~~    \\
& \\
S$^{+}$/H$^{+}$($\times$10$^7$)\     & 2.32$\pm$0.36~~    &  1.76$\pm$0.33~~   & 1.80$\pm$0.49~~      \\
S$^{++}$/H$^{+}$($\times$10$^7$)\    & 4.10$\pm$7.71~~    &  9.00$\pm$4.12~~   & \nodata              \\
ICF(S)\                              & 1.22               &  1.19              & 1.22                 \\
S/H($\times$10$^7$)\                 & 7.85$\pm$9.45~~    & 12.75$\pm$4.90~~   & \nodata              \\
12+log(S/H)\                         & 5.90$\pm$0.52~~    &  6.11$\pm$0.17~~   & \nodata              \\
log(S/O)\                            & --1.65$\pm$0.53~~  &  --1.43$\pm$0.19~~ & \nodata              \\
& \\
Ar$^{++}$/H$^{+}$($\times$10$^7$)\   & 2.13$\pm$0.50~~    &  1.81$\pm$0.57~~   & 2.15$\pm$0.51~~      \\
ICF(Ar)\                             & 1.44               &  1.48              & 1.45                 \\
Ar/H($\times$10$^7$)\                & 3.08$\pm$0.73~~    &  2.69$\pm$0.84~~   & 3.11$\pm$0.74~~      \\
12+log(Ar/H)\                        & 5.49$\pm$0.10~~    &  5.43$\pm$0.14~~   & 5.49$\pm$0.10~~      \\
log(Ar/O)\                           & --2.05$\pm$0.14~~  &  --2.11$\pm$0.17~~ & --2.03$\pm$0.15~~    \\
\enddata
\end{deluxetable}

%
%
\begin{deluxetable}{lccc}
\tabletypesize{\footnotesize}
\tablecaption{Abundances in Sextans B H\,{\sc ii} regions
	      \label{tbl:SextB_HII_ab}}
\tablewidth{0pt}
\tablehead{
\colhead{Value}    & \colhead{H\,{\sc ii}$-$5} & \colhead{H\,{\sc ii}$-$1} & \colhead{H\,{\sc ii}$-$2} \\
\colhead{(1)}      & \colhead{(2)}   & \colhead{(3)}   & \colhead{(4)}
}
\startdata
$T_{\rm e}$(OIII)(K)\                & 12,760$\pm$640~~   & 17,230$\pm$1620~~  & 17,450$\pm$1160~~   \\
$T_{\rm e}$(OII)(K)\                 & 12,590$\pm$590~~   & 14,730$\pm$1320~~  & 14,820$\pm$940~~    \\
$T_{\rm e}$(SIII)(K)\                & 12,680$\pm$530~~   & 16,000$\pm$1340~~  & 16,180$\pm$960~~    \\
$N_{\rm e}$(SII)(cm$^{-3}$)\         &  10$\pm$10~~       &  10$\pm$10~~       &  10$\pm$10~~        \\
& \\
O$^{+}$/H$^{+}$($\times$10$^5$)\     & 2.82$\pm$0.49~~    & 1.85$\pm$0.54~~    & 2.41$\pm$0.49~~     \\
O$^{++}$/H$^{+}$($\times$10$^5$)\    & 4.12$\pm$0.59~~    & 1.31$\pm$0.30~~    & 1.12$\pm$0.18~~     \\
O/H($\times$10$^5$)\                 & 6.94$\pm$0.77~~    & 3.16$\pm$0.62~~    & 3.52$\pm$0.53~~     \\
12+log(O/H)\                         & ~7.84$\pm$0.05~~   & ~7.50$\pm$0.08~~   & ~7.55$\pm$0.06~~    \\
& \\
N$^{+}$/H$^{+}$($\times$10$^7$)\     & 9.72$\pm$0.95~~    & 5.83$\pm$1.03~~    & 6.82$\pm$0.89~~     \\
ICF(N)\                              & 2.46               & 1.71               & 1.46                \\
N/H($\times$10$^7$)\                 & 23.91$\pm$2.34~    & 9.95$\pm$1.75~~    & 9.98$\pm$1.30~~     \\
12+log(N/H)\                         & 6.38$\pm$0.04~~    & 6.00$\pm$0.08~~    & 6.00$\pm$0.06~~     \\
log(N/O)\                            & --1.46$\pm$0.06~~  & --1.50$\pm$0.11~~  & --1.55$\pm$0.09~~   \\
& \\ 
Ne$^{++}$/H$^{+}$($\times$10$^5$)\   & 0.55$\pm$0.09~~    & \nodata            & \nodata             \\
ICF(Ne)\                             & 1.69               & \nodata            & \nodata             \\
Ne/H($\times$10$^5$)\                & 0.93$\pm$0.15~~    & \nodata            & \nodata             \\
12+log(Ne/H)\                        & 6.97$\pm$0.07~~    & \nodata            & \nodata             \\
log(Ne/O)\                           & --0.87$\pm$0.08~~  & \nodata            & \nodata             \\
& \\ 
S$^{+}$/H$^{+}$($\times$10$^7$)\     & 3.661$\pm$0.304~~  & 2.422$\pm$0.339~~  & 2.835$\pm$0.277~~   \\
S$^{++}$/H$^{+}$($\times$10$^7$)\    & 19.220$\pm$2.998~~ & 5.271$\pm$1.655~~  & 4.740$\pm$1.670~~   \\
ICF(S)\                              & 1.252              & 1.158              & 1.100               \\
S/H($\times$10$^7$)\                 & 28.64$\pm$3.77~    & 8.91$\pm$1.96~~    & 8.33$\pm$1.86~~     \\
12+log(S/H)\                         & 6.46$\pm$0.07~~    & 5.95$\pm$0.10~~    & 5.92$\pm$0.10~~     \\
log(S/O)\                            & --1.38$\pm$0.07~~  & --1.55$\pm$0.13~~  & --1.63$\pm$0.12~~   \\
& \\
Ar$^{++}$/H$^{+}$($\times$10$^7$)\   & 3.55$\pm$0.28~~    & 1.80$\pm$0.26~~    & 1.77$\pm$0.22~~     \\
Ar$^{+++}$/H$^{+}$($\times$10$^7$)\  & 3.78$\pm$0.70~~    & 0.00$\pm$0.00~~    & 0.00$\pm$0.00~~     \\
ICF(Ar)\                             & 1.18               & 1.55               & 1.82                \\
Ar/H($\times$10$^7$)\                & 8.64$\pm$0.89~~    & 2.79$\pm$0.40~~    & 3.22$\pm$0.40~~     \\
12+log(Ar/H)\                        & 5.94$\pm$0.04~~    & 5.45$\pm$0.06~~    & 5.50$\pm$0.05~~     \\
log(Ar/O)\                           & --1.90$\pm$0.07~~  & --2.05$\pm$0.11~~  & --2.04$\pm$0.08~~   \\
\enddata
\end{deluxetable}

%
%
\begin{deluxetable}{lcccc}
\tablecaption{Line intensities of observed PNe in Sextans A and Sextans B
	      \label{tbl:SextAB_PN_int}}
\tablewidth{0pt}
\tablehead{
		    & \MC{2}{c}{PN in Sextans A}                                              & \MC{2}{c}{PN3 in Sextans B}                                             \\  \hline
\colhead{(\AA) Ion} & \colhead{F($\lambda$)/F(H$\beta$)} & \colhead{I($\lambda$)/I(H$\beta$)} & \colhead{F($\lambda$)/F(H$\beta$)} & \colhead{I($\lambda$)/I(H$\beta$)} \\
\colhead{(1)}       & \colhead{(2)}                      & \colhead{(3)}                      & \colhead{(4)}                      & \colhead{(5)}
}
\startdata
4101\ H$\delta$\                          & 0.163$\pm$0.013 & 0.191$\pm$0.021 & 0.242$\pm$0.035 & 0.271$\pm$0.050  \\
4340\ H$\gamma$\                          & 0.438$\pm$0.017 & 0.475$\pm$0.022 & 0.386$\pm$0.059 & 0.416$\pm$0.068  \\
4363\ [O\ {\sc iii}]\                     & 0.182$\pm$0.014 & 0.193$\pm$0.015 & 0.163$\pm$0.047 & 0.175$\pm$0.051  \\
4686\ He\ {\sc ii}\                       & 0.481$\pm$0.016 & 0.489$\pm$0.017 & \nodata         & \nodata          \\
4740\ [Ar\ {\sc iv]}\                     & 0.007$\pm$0.005 & 0.007$\pm$0.005 & \nodata         & \nodata          \\
4861\ H$\beta$\                           & 1.000$\pm$0.024 & 1.000$\pm$0.025 & 1.000$\pm$0.072 & 1.000$\pm$0.074  \\
4959\ [O\ {\sc iii}]\                     & 2.160$\pm$0.043 & 2.121$\pm$0.042 & 1.377$\pm$0.087 & 1.359$\pm$0.086  \\
5007\ [O\ {\sc iii}]\                     & 6.340$\pm$0.111 & 6.190$\pm$0.110 & 4.177$\pm$0.222 & 4.096$\pm$0.219  \\
5199\ [N\ {\sc i}]\                       & 0.475$\pm$0.016 & 0.454$\pm$0.016 & \nodata         & \nodata          \\
5755\ [N\ {\sc ii}]\                      & 0.287$\pm$0.016 & 0.258$\pm$0.014 & \nodata         & \nodata          \\
5876\ He\ {\sc i}\                        & 0.115$\pm$0.009 & 0.102$\pm$0.008 & 0.195$\pm$0.038 & 0.173$\pm$0.034  \\
6300\ [O\ {\sc i}]\                       & 0.808$\pm$0.020 & 0.691$\pm$0.018 & \nodata         & \nodata          \\
6312\ [S\ {\sc iii}]\                     & 0.007$\pm$0.004 & 0.006$\pm$0.004 & \nodata         & \nodata          \\
6364\ [O\ {\sc i}]\                       & 0.295$\pm$0.014 & 0.251$\pm$0.012 & \nodata         & \nodata          \\
6548\ [N\ {\sc ii}]\                      & 3.091$\pm$0.062 & 2.589$\pm$0.057 & \nodata         & \nodata          \\
6563\ H$\alpha$\                          & 3.266$\pm$0.064 & 2.757$\pm$0.066 & 3.074$\pm$0.170 & 2.601$\pm$0.153  \\
6584\ [N\ {\sc ii}]\                      & 9.587$\pm$0.166 & 8.007$\pm$0.156 & 0.074$\pm$0.028 & 0.061$\pm$0.023  \\
6678\ He\ {\sc i}\                        & 0.035$\pm$0.007 & 0.029$\pm$0.006 & \nodata         & \nodata          \\
6717\ [S\ {\sc ii}]\                      & 0.055$\pm$0.008 & 0.046$\pm$0.007 & \nodata         & \nodata          \\
6731\ [S\ {\sc ii}]\                      & 0.074$\pm$0.008 & 0.061$\pm$0.007 & \nodata         & \nodata          \\
7065\ He\ {\sc i}\                        & 0.034$\pm$0.011 & 0.027$\pm$0.009 & 0.177$\pm$0.034 & 0.140$\pm$0.031  \\
7136\ [Ar\ {\sc iii}]\                    & 0.015$\pm$0.007 & 0.012$\pm$0.006 & \nodata         & \nodata          \\
7320\ [O\ {\sc ii}]\                      & 0.154$\pm$0.019 & 0.121$\pm$0.015 & 0.142$\pm$0.070 & 0.112$\pm$0.055  \\
7330\ [O\ {\sc ii}]\                      & 0.119$\pm$0.019 & 0.094$\pm$0.015 & \nodata         & \nodata          \\
& & & & \\
C(H$\beta$)\ dex                          & \MC{2}{c}{0.23$\pm$0.03}          & \MC{2}{c}{0.25$\pm$0.07} \\
F(H$\beta$)\tablenotemark{a}              & \MC{2}{c}{4.9$\pm$0.1}           & \MC{2}{c}{2.2$\pm$0.1} \\
\enddata
\tablenotetext{a}{Observed flux in units of 10$^{-16}$ ergs\ s$^{-1}$cm$^{-2}$}
\end{deluxetable}

%
%
\begin{deluxetable}{lcc}
\tabletypesize{\footnotesize}
\tablecaption{Abundances of observed PNe in Sextans A and Sextans B
	      \label{tbl:SextAB_PN_ab}}
\tablewidth{0pt}
\tablehead{
\colhead{Value}    & \colhead{PN in Sextans A} & \colhead{PN3 in Sextans B} \\
\colhead{(1)}      & \colhead{(2)}             & \colhead{(3)}
}
\startdata
$T_{\rm e}$(OIII)(K)$^a$\            & 19,020$\pm$860 ~~      & 23,300$\pm$4860~~   \\
$T_{\rm e}$(OII)(K)\                 & 15,380$\pm$650 ~~      & 16,450$\pm$2770~~   \\
$T_{\rm e}$(NII)(K)$^a$\             & 13,750$\pm$530 ~~      & \nodata             \\
$T_{\rm e}$(SIII)(K)\                & 17,490$\pm$710 ~~      & \nodata             \\
$N_{\rm e}$(SII)(cm$^{-3}$)\         & 1918$\pm$1434~~        & 2000                \\
& &\\
O$^{+}$/H$^{+}$($\times$10$^5$)\     & 2.95$\pm$0.40~~        & 1.17$\pm$0.74~~     \\
O$^{++}$/H$^{+}$($\times$10$^5$)\    & 3.96$\pm$0.42~~        & 1.76$\pm$0.79~~     \\
O$^{+++}$/H$^{+}$($\times$10$^5$)\   & 3.54$\pm$0.77~~        & \nodata             \\
O/H($\times$10$^5$)\                 & 10.45$\pm$0.96~~       & 2.93$\pm$1.08~~     \\
12+log(O/H)\                         & ~8.02$\pm$0.05~~       & ~7.47$\pm$0.16~~    \\
& &\\
N$^{+}$/H$^{+}$($\times$10$^6$)\     & 71.60$\pm$5.36~        & 0.39$\pm$0.16~~     \\
ICF(N)\                              & 3.54                   & 2.51                \\
N/H($\times$10$^5$)\                 & 25.37$\pm$1.90~~       & 9.81$\pm$3.97~~     \\
12+log(N/H)\                         & 8.40$\pm$0.03~~        & 5.99$\pm$0.18~~     \\
log(N/O)\                            & 0.39$\pm$0.05~~        & --1.48$\pm$0.24~~   \\
& &\\
S$^{+}$/H$^{+}$($\times$10$^7$)\     & 1.23$\pm$0.21~~        & \nodata             \\
S$^{++}$/H$^{+}$($\times$10$^7$)\    & 1.96$\pm$1.18~~        & \nodata             \\
ICF(S)\                              & 1.38                   & \nodata             \\
S/H($\times$10$^7$)\                 & 4.38$\pm$1.64~~        & \nodata             \\
12+log(S/H)\                         & 5.64$\pm$0.16~~        & \nodata             \\
log(S/O)\                            & --2.38$\pm$0.17~~      & \nodata             \\
& &\\
Ar$^{++}$/H$^{+}$($\times$10$^7$)\   & 0.35$\pm$0.17~~        & \nodata             \\
Ar$^{+++}$/H$^{+}$($\times$10$^7$)\  & 0.64$\pm$0.45~~        & \nodata             \\
ICF(Ar)\                             & 1.08                   & \nodata             \\
Ar/H($\times$10$^7$)\                & 1.07$\pm$0.52~~        & \nodata             \\
12+log(Ar/H)\                        & 5.03$\pm$0.21~~        & \nodata             \\
log(Ar/O)\                           & --2.99$\pm$0.21~~      & \nodata             \\
& &\\
He/H\                                & 0.11$\pm$0.01~~        & \nodata             \\
12+log(He/H)\                        & 11.03$\pm$0.03~~       & \nodata             \\
& \\ 
\enddata
\tablenotetext{a}{Temperature was calculated directly}
\end{deluxetable}

%
%
\begin{deluxetable}{lcc}
\tablecaption{Physical Parameters of observed PNe in Sextans A and Sextans B
	      \label{tbl:PN_par}}
\tablewidth{0pt}
\tablehead{
\colhead{Property} & \colhead{Sextans A PN} & \colhead{Sextans B PN3} \\
\colhead{(1)}      & \colhead{(2)}          & \colhead{(3)}
}
\startdata
T$_{eff}$ (K)      & 180,000                & $\le$95,000         \\
log10(T$_{eff}$)   & 5.26                   & $\le$4.98           \\
L/L$_\odot$        & 6500                   & 3400                \\
log(L/L$_\odot$)   & 3.80                   & 3.53                \\
M/M$_\odot^1$      & $\sim$1.5$^a$          & $\le$1.0$^a$        \\
t$_{MS}$ (Gyr)$^2$ & $\sim$1.6$^a$          & $\ge$5.7$^a$        \\
		   &                        &                     \\
V (mag)$^b$        & 22\fm41$\pm$0\fm16     & 22\fm71$\pm$0\fm27  \\
R (mag)$^b$        & 21\fm57$\pm$0\fm12     & 22\fm51$\pm$0\fm38  \\
I (mag)$^b$        & 23\fm43$\pm$0\fm29     & 22\fm40$\pm$0\fm64  \\
A$_V$ (mag)$^c$    & 0\fm50                 & 0\fm26              \\
\enddata
\tablerefs{(1) \citet{VW94};
	   (2) \citet{VW93}
}
\tablenotetext{a}{For Z=0.001 (1/20 of Z$_{\odot}$)}
\tablenotetext{b}{Total observed magnitudes calculated on the base of spectra
as described in Section~\ref{txt:Chem_PN} and non-corrected for the extinction}
\tablenotetext{c}{Total extinction in the V band calculated from $C$(H$\beta$) using
equation A$_V = 3.2 \cdot E_{B-V} = 3.2 \cdot 0.68 \cdot C(H\beta)$}
\end{deluxetable}

\end{document}